%% file: main.tex
\documentclass[final,conference,10pt]{IEEEtran}
\usepackage{color,overpic,cite}
\usepackage{pdfsync}
\usepackage{url}
\usepackage{enumitem}
\usepackage{setspace}
\usepackage{amsmath,amsfonts,amssymb,amsthm,bbm,bm,comment,multirow,subfigure}
\usepackage[ruled,commentsnumbered]{algorithm2e} 
\usepackage{tcolorbox}

\newcommand{\eat}[1]{{}}
\newcommand{\mean}[1]{\mathbb{E}\!\left[#1\right]}
\newcommand{\prob}[1]{\mathbb{P}\!\left(#1\right)}
\newcommand{\argmin}{\arg\min}

\newtheorem{theorem}{Theorem}
\newtheorem{lemma}{Lemma}

\newtheorem{definition}{Definition}

\newtheorem{corollary}{Corollary}

\newcommand{\ev}{{\bf e}}

\newcommand{\Yc}{{\cal Y}}

\newcommand{\Nc}{{\cal N}}

\newcommand{\Mc}{{\cal M}}

\newcommand{\Sigmam}{\hbox{\boldmath$\Sigma$}}

\newcommand{\rev}[1]{{\color{black}#1}} 

\title{An OCO Approach To Optimizing Caching Networks}
\title{Online Convex Optimization for Caching Networks}
\author{Georgios S. Paschos, Apostolos Destounis, and George Iosifidis\thanks{{ Georgios S. Paschos is with Amazon, Luxembourg  (e-mail: gpaschos@gmail.com). Apostolos Destounis is with the France Research Center, Huawei Technologies, 92100 Boulogne-Billancourt, France (e-mail: apostolos.destounis@huawei.com). George Iosifidis is with the School of Computer Science and Statistics, Trinity College Dublin, The University of Dublin, College Green, Dublin 2, D02PN40 Ireland (e-mail: george.iosifidis@tcd.ie). Part of this work has appeared in the proceedings of IEEE INFOCOM 2019 \cite{paschos-infocom19}. The work of G. Iosifidis is supported by the Science Foundation Ireland under Grants 17/CDA/4760 and 16/IA/4610.}}}

\IEEEoverridecommandlockouts
\begin{document}
\maketitle
\pagestyle{plain}

\begin{abstract}

We study the problem of wireless edge caching when file popularity is unknown and possibly non-stationary. A bank of $J$ caches receives file requests and a utility is accrued for each request depending on the serving cache. The network decides dynamically which files to store at each cache and how to route them, in order to maximize total utility. The request sequence is assumed to be drawn from an arbitrary distribution, capturing time-variance, temporal and spatial locality of requests. For this challenging setting, we propose the \emph{Bipartite Supergradient Caching Algorithm} (BSCA) which provably exhibits no regret ($R_T/T \to 0$). That is, as the time horizon $T$ increases, BSCA achieves (at least) the same utility with the cache configuration that we would have chosen knowing all future requests. The learning rate of the algorithm is characterized by its regret expression $R_T\!=\!O(\sqrt{JT})$, which is independent of the file library size. For the single-cache case, we prove that this is the lowest attainable bound. BSCA requires at each step $J$ projections on intersections of boxes and simplices, for which we propose a tailored algorithm. Our model is the first that draws a connection between the network caching problem and \emph{Online Convex Optimization}, and we demonstrate its generality by discussing various practical extensions and presenting a trace-driven comparison with state-of-the-art competitors. 

\end{abstract}
\input{introduction_R1.tex}
\input{related_works_R1.tex}

\input{system_model_R1.tex}
\input{bipartite_R1.tex}
\input{projection_R1.tex}

\input{gradient_descent_R1.tex}

\input{extensions_R1.tex}

\input{simulations_R1.tex}

\input{conclusions_R1.tex}
\bibliography{mybib_v13}
\bibliographystyle{IEEEtran}

\begin{IEEEbiography}[{\includegraphics[width=1.25in,height=1.25in,clip,keepaspectratio]{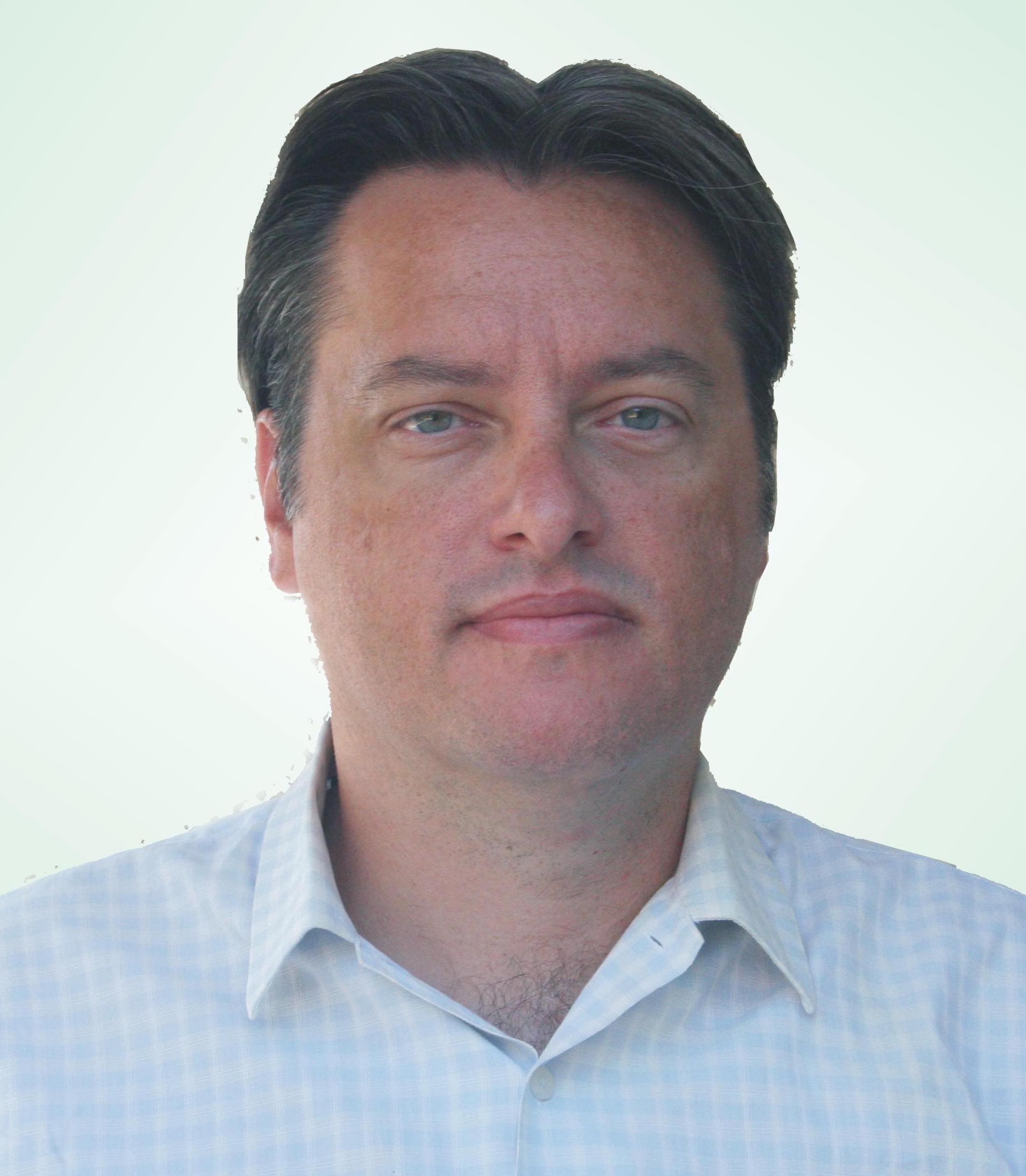}}]
	{Georgios Paschos} is is a Senior Manager, Research Science at Amazon.com, leading the EU Operation Research team of Amazon Transportation Services. Previously, he worked 5 years ('14-'19) as a principal scientist at Huawei Technologies, Paris, leading the Network Control and Resource Allocation team. Dr. Paschos was at LIDS, MIT ('12-'14) and has held positions at CERTH-ITI, Greece '08-'12 (researcher), University of Thessaly, '09-'11 (adjunct lecturer) and VTT, Finland, '07-'08 (ERCIM Postdoc Fellow). He received his diploma in Electrical and Computer Engineering in 2002 from Aristotle University of Thessaloniki, and his PhD degree in Wireless Networks 2006 from ECE dept. University of Patras, Greece. Two of his papers won best paper awards in GLOBECOM 2007 and IFIP Wireless Days 2009. He has served as an associate editor for IEEE/ACM Trans. on Networking ('15-'19), IEEE Networking Letters ('18-'19), and as a TPC member of INFOCOM, WiOPT, and Netsoft. He has organized several international workshops on the topics of caching, network slicing and machine learning techniques for communication systems, and was the co-organizer and editor of the IEEE JSAC Special Issue on Caching for Comm. Systems and Networks.
\end{IEEEbiography}

\begin{IEEEbiography}[{\includegraphics[width=1.121in,height=1.25in,clip]{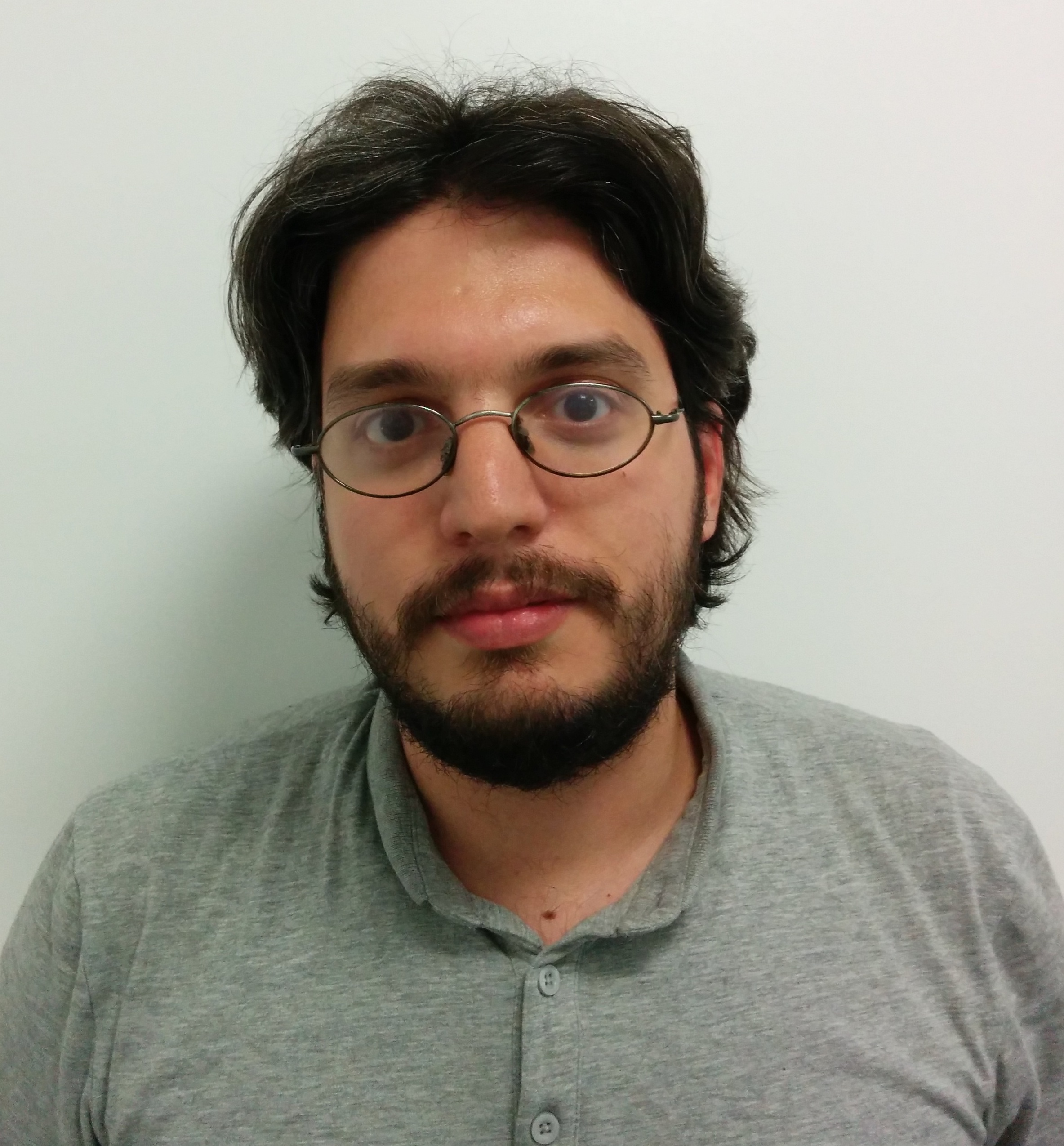}}]
	{Apostolos Destounis} received the Diploma in Electrical and Computer Engineering from the National technical University of Athens, the M.Sc. in Communications and Signal Processing from Imperial College London and the Ph.D. In Telecommunications from CentraleSupelec, Paris in 2009, 2010 and 2014, respectively. From 2011 to 2014 he also was a Research Engineer in Alcetel-Lucent (now Nokia) Bell Labs France. Since May 2014 he has been with the Mathematical and Algorithmic Sciences Lab, Paris Research Center, Huawei Technologies co. Ltd, where he is a Senior Research Engineer. He was the co-organizer of the first International Workshop on Machine Learning for Communications (WMLC), hosted with WiOpt 2019. His current research interests lie in the fields of optimization and machine learning, with applications to wireless networks and content caching. 
\end{IEEEbiography}

\begin{IEEEbiography}[{\includegraphics[width=1.25in,height=1.25in,clip,keepaspectratio]{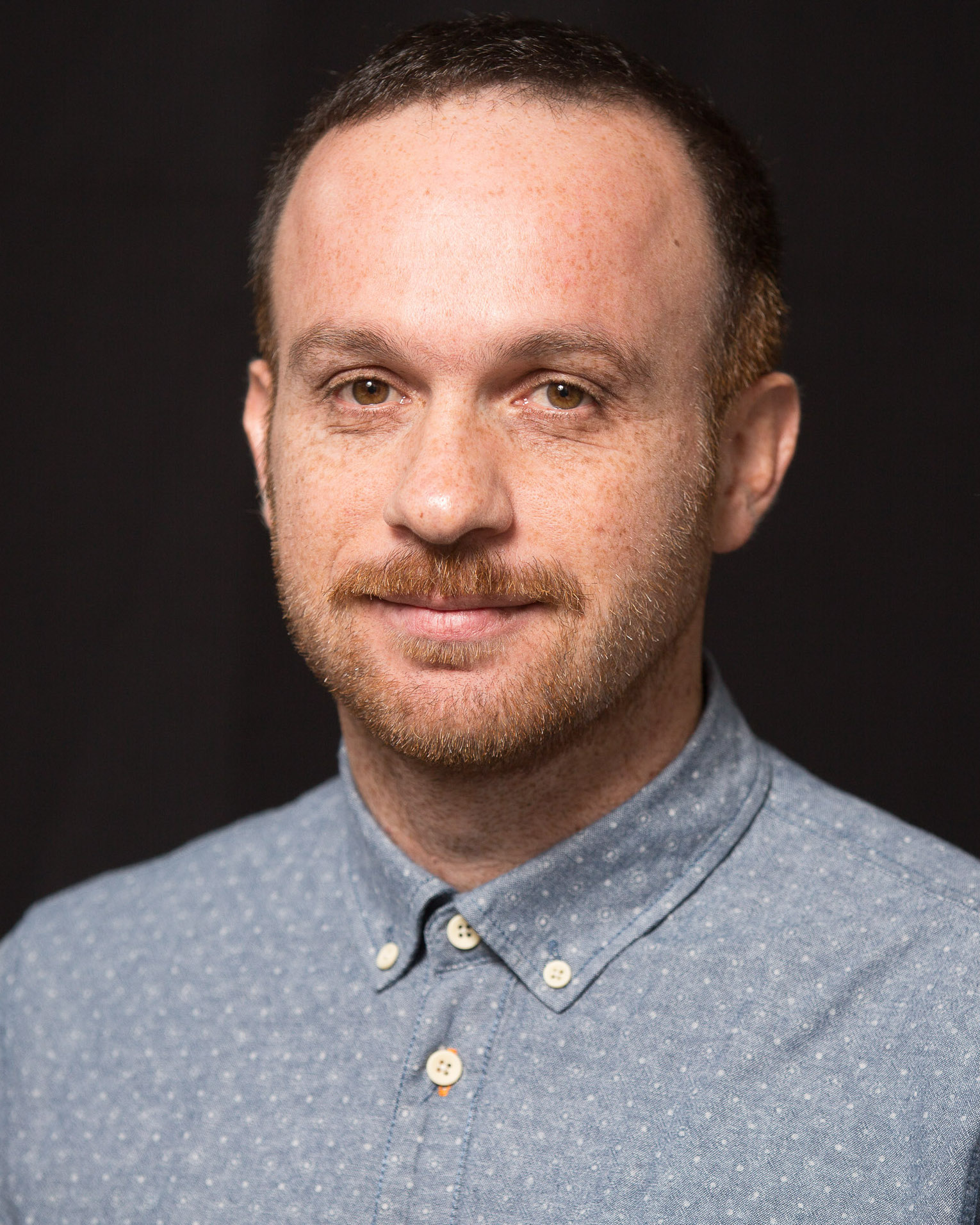}}]
	{George Iosifidis} is an Assistant Professor with Trinity College Dublin, Ireland. He received a Diploma in Electronics and Telecommunications Engineering from the Greek Air Force Academy (Athens, 2000), and a PhD degree from the ECE Dep., University of Thessaly in 2012. He was a Postdoctoral researcher ('12-'14) at CERTH, and at Yale University ('14-'17). His research interests lie in the broad area of network optimization and economics, and his work has appeared in Nature Communications, Nature Human Behavior and PNAS. He is a co-recipient of the best paper awards in IEEE WiOPT 2013 and IEEE INFOCOM 2017, has served as a guest editor for the IEEE Journal on Selected Areas in Communications, and is currently an editor for IEEE Transactions on Communications and IEEE/ACM Transactions on Networking. 
\end{IEEEbiography}

\end{document}

%% file: introduction_R1.tex
\section{Introduction}


The wireless edge caching architecture proposes to cache popular files at small-cell base stations (SBS) in order to serve future user requests \cite{femtocaching}. This is a promising approach for accommodating the increasing mobile data traffic in a cost-efficient fashion \cite{paschos_16}, and has rightfully spurred a flurry of related work \cite{paschos-jsac}. A weakness of these proactive caching solutions, however, is that they assume static and known file popularity. Practice has shown quite the opposite: \emph{file popularity changes fast, and it is challenging to learn it}. Here, we study these systems from a new perspective and \emph{propose an online caching policy that optimizes their performance under any popularity model}. Our approach tackles the caching problem in its most general form and reveals a novel connection between (wireless or wired) caching networks and Online Convex Optimization (OCO) \cite{zinkevich2003online}.

\subsection{Motivation}

Due to its finite capacity a cache can host only a small subset of the file library, and it is therefore necessary to employ a \emph{caching policy} that selects which files should be stored. The main selection criterion is typically the fraction of file requests the cache can satisfy (\emph{cache hit ratio}), and different policies employ different rules in order to maximize this metric. For instance, the Least-Recently-Used (LRU) policy inserts in the cache the newly requested file and evicts the one that has not been requested for the longest time period; while the Least-Frequently-Used (LFU) policy evicts the file that is least frequently requested. These widely adopted policies were designed empirically, and hence a question that arises is \emph{under what conditions they achieve high hit ratio}?


The answer depends on the file popularity model. For instance, it has been shown that \emph{(i)} for stationary requests, LFU achieves the highest hit ratio~\cite{Fricker12}; \emph{(ii)} an \emph{age-based-threshold} policy maximizes the ratio when requests follow the Poisson Shot Noise model \cite{snm}; and \emph{(iii)} LRU has the highest hit ratio \cite{Sleator85} for more general request models \cite{Belady66, Mattson70}. These policies, however, perform poorly when the request model is other than the one assumed \cite{paschos-infocom19}; and indeed in practice the requests follow unknown and possibly time-varying distributions. \emph{This renders imperative the design of a {universal} caching policy that works provably well for all possible request models.}

\begin{figure}[t!]
	\centering
	\subfigure{\includegraphics[width=3.55in]{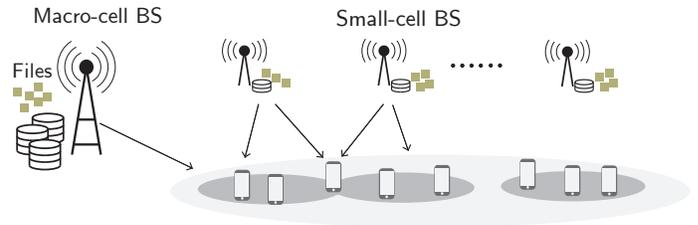}\label{Fig:intro1}}
	\caption{\textbf{Wireless Edge Caching}. Storage-endowed SBSs underlay the main base station (MBS) and can serve user requests with cached content over short-range links. The MBS can serve all users and has direct access to the entire file library, while each SBS can only store a small portion of it.}
	\label{Fig:motivation}
\end{figure}

This requirement is even more crucial for wireless edge caching networks, see Fig. \ref{Fig:motivation}, where the caches receive requests with low rate and therefore ``see'' processes with highly non-stationary behavior \cite{mathieu, Elayoubi2015}. Moreover, due to the wireless medium, a user might be within the range of multiple SBS caches, each one offering a different transmission rate and thus \emph{caching utility}. This creates the need for explicit routing decisions which are inevitably intertwined with the caching policy. In other words, the caching decisions across different SBSs are coupled, routing affects caching, and the requests might change both in space and time. Request models for this intricate case include random replacement models \cite{Elayoubi2015}, and inhomogeneous Poisson processes \cite{snm, kauffman}, among others. However, such multi-parametric models are challenging to fit to data, and rely on strong assumptions about the popularity evolution (see Sec. \ref{sec:related}). Our approach is orthogonal to these works as we design an online learning mechanism for adaptive caching and routing decisions that reduce the MBS transmissions and maximize the utility offered by the SBS caches.

\subsection{Methodology and Contributions}

We introduce a model-free caching model along the lines of the OCO framework. We assume that file requests are drawn from a general distribution, which is equivalent to caching versus an \emph{adversary} \rev{that chooses the requests arbitrarily.}\footnote{\rev{The adversary might even select the requests attempting to degrade the system performance, exploit our past caching decisions, and so on.}} At each slot \emph{(i)} the adversary creates a new file request; \emph{(ii)} a routing plan is deployed to retrieve the file from the SBS caches and/or MBS; \emph{(iii)} a (file, cache)-dependent utility is obtained; and \emph{(iv)} the caching policy updates the stored files at each SBS. This generalizes the criterion of cache hit ratio and allows one to build policies that, for instance, minimize delay or provide different priority to different users.

%

In this setting we seek to design a policy with \emph{sublinear regret}; i.e., one that achieves zero utility loss per slot as the time horizon $T$ increases when it is compared to the best static cache configuration (\emph{hindsight policy}). To this end, we propose the \emph{Bipartite Supergradient Caching Algorithm} (BSCA) policy, and prove that its regret $R_T$ is upper bounded by $w^{(1)}\sqrt{2\text{deg} JCT}$  for a network of $J$ caches that each can store up to $C$ of the $N$ library files. Constants $w^{(1)}$ and deg are independent of parameters $T$ and $N$; and therefore BSCA amortizes the average loss compared to the hindsight policy, i.e. $R_T/T\to 0$, and its oblivious to the library size. Moreover, for the single cache scenario we derive the lower attainable regret bound and prove that BSCA matches it. Our contributions can be thus summarized as follows:

\begin{itemize}[leftmargin=4mm]
	\item \emph{Machine Learning (ML) caching}: We provide a fresh ML angle for the design of wireless edge caching policies by reformulating this problem to handle time-varying file popularity and ensure its efficient solution. To the best of our knowledge this is the first time online convex optimization is used in the context of caching networks.
	
	\item  \emph{Universal caching policy}: BSCA has zero loss over the hindsight policy under any request model and hits the sweet spot of complexity versus performance. It is applicable to a variety of settings, including general caching networks that can be modeled with a bipartite graph, and networks with time-varying parameters or file prefetching costs.
	
	\item \emph{Single-cache performance}: For the basic model of one cache, we prove that the lowest attainable regret is $R_T=\Omega(B\sqrt{CT})$, where parameter $B$ is independent of $T$. We show that BSCA achieves this bound, by employing a smart combination of LFU and LRU-type  decisions.

	\item \emph{Fast Cache Projection}: BSCA requires at each slot $J$ projections on the intersection of box and simplex constraints. We design a routine that performs each of them in $O(N)$ steps. This simplifies the execution of BSCA and enables its application to large caching networks.

	\item \emph{Trace-driven Evaluation}: We evaluate BSCA using several request models and real traces, and compare it with state-of-the-art competitor algorithms. We verify that BSCA has no regret and we find that it outperforms previous policies by up to 45.8$\%$ in typical scenarios.
\end{itemize}

\subsection{Paper Organization} 

The rest of this paper is organized as follows. Section \ref{sec:related} presents the related work and Section \ref{sec:model} introduces the system model. The online wireless edge caching problem is formulated in Section \ref{sec:formulation}, and Section \ref{sec:bipartite} presents the BSCA algorithm for a network of caches. Section \ref{sec:projection} introduces our projection routine. Section \ref{sec:gradient} focuses on the simpler but important case of one cache. We discuss model extensions in Section \ref{sec:extensions}, compare BSCA with key competitors in Section \ref{sec:performance} and conclude in Section \ref{sec:conclusions}.

%% file: related_works_R1.tex
\section{Background and Related Work}\label{sec:related}

The literature of caching policies cannot, by any means, be covered in a single section, and we refer the interested reader to \cite{paschos-jsac, paschos-fnt} for a thorough presentation. We focus here on reactive policies and online algorithms for caching networks.

\subsection{Reactive Policies}

The design of caching policies depends heavily on the file popularity model that is assumed to generate the requests. One option is to use the adversarial model of \cite{Sleator85}, where a policy's hit rate is compared to Belady's dynamic hindsight policy that evicts the file which will be requested farther in the future \cite{Belady66, Mattson70}. LRU performs better than other policies under this model\footnote{Comparing with this very demanding benchmark requires one to restrict the caching cacpacity of the Belady cache to a portion of the actual cache --- otherwise all policies perform very poorly.} \cite{Sleator85}, but its performance is actually comparable to any other \emph{marking} policy \cite{Borodin98}, e.g., even to a simple FIFO. In a sense, this dynamic hindsight policy is a ``too strong'' benchmark to help us identify a good caching policy. On the other hand, moderately stationary models like IRM \cite{Fricker12} are easy to fit in data, and LFU maximizes the cache hit rate in this case. However, IRM is accurate only when used to model requests within small time intervals where popularity is roughly static, hence it is not suitable for evaluating long-term performance of a caching network.

In fact, in real systems the requests are rarely stationary and this has motivated the proposal of several non-stationary models. For instance, \cite{Gitz_14} uses the theory of variations, \cite{Elayoubi2015} makes random content replacements in the catalog, \cite{kauffman} proposes a time-dependent Poisson model, and \cite{snm} introduced the shot noise model for correlated requests in temporal proximity. Unfortunately, selecting and fitting these models to data is a time-demanding task \cite{mathieu}, and thus not suitable for fast-changing environments. There are also several model-based/free approaches for predicting content popularity using statistical analysis, transfer learning, or social network properties, see \cite{szabo10, Bastug2015TransferExtended, poor-learning-caching, schaar-forecasting-jsasp}. Yet, these works do not incorporate the predictions into the system operation. Unlike prior efforts, our proposal does not involve model selection and \emph{the learning mechanism is fully embedded into the caching policy}.

Instead of fitting models, another  option is to learn the popularity without using prior assumptions \cite{gunduz-reinforcement, giannakis-q-learning}. For instance, \cite{giannakis-q-learning} models the popularity evolution as a Markov process and employs \emph{Q-learning} to estimate the transition probabilities which are then used for proactive caching. Such model-free solutions work well if there are adequate data, but have substantial computation and memory requirements. For instance, tabular Q-learning needs memory size combinatorial in the catalog size and cache capacity; and Q-learning with function approximation requires more involved gradient computations, while its convergence can be slow. Following a different approach, \cite{mihaela-video-caching} predicts file popularities using classification. This interesting approach, however, needs feature extraction, does not consider routing, nor accounts for changes in utilities. Other online caching proposals include \cite{geulen2010regret, englert2013economical, lykouris-ML} which study the basic paging problem of hit-maximization in one cache. \emph{Our approach works for networks of caches without requiring stationary or known request models.}

\subsection{Caching Networks (CNs)} 

The first OCO-based caching policy was proposed in \cite{paschos-infocom19} which reformulated the caching problem and embedded a learning mechanism, while \cite{paschos-spawc} studied how such policies can be used in device-to-device caching scenarios. In CNs one needs to additionally decide which cache will satisfy a request (routing) and which files will be evicted (caching), and these decisions are perplexed when each user is connected to multiple caches. Thus, it is not surprising that online policies for CNs are under-explored. Placing more emphasis on the network, \cite{ioannidis-jsac18} introduced a joint routing and caching algorithm assuming that file popularity is stationary. On the other hand, proposals for reactive CN policies include: randomized caching policies for small-cell networks \cite{Blaszczyszyn2014Geographic}; joint caching and SBSs transmission policies \cite{spyropoulos-icc}; distributed cooperative caching algorithms \cite{avrachenkov-acm17}; and policies using a TTL-based utility-cache model \cite{C_Dehghan_16}. All these solutions \emph{presume that the popularity model is fixed and known}.

Finally, \cite{giovanidis-mLRU} proposed the multi-LRU (mLRU) heuristic strategy, and \cite{leonardi-implicit} the ``lazy rule'' extending $q$-LRU to provide local optimality guarantees under stationary requests. These works pioneered the extension of the seminal LFU/LRU-type policies to the case of multiple connected caches and designed efficient caching algorithms with minimal overheads. Nevertheless, dropping the stationarity assumption, the problem of online routing and caching remains open. Our method is different as we embed a learning mechanism into the system operation that \emph{adapts the caching and routing decisions to any request model and to network changes}.

%% file: system_model_R1.tex
\section{System model}\label{sec:model}
\textbf{Network Connectivity}. The caching network consists of small-cell base stations (SBS) denoted with the set ${\cal J}\!=\!\{1,2,\dots, J\}$, and a macro-cell base station (MBS) indexed with 0; each station is equipped with a cache. There is  a set of user locations ${\cal I}=\{1,2,\dots, I\}$, where file requests are created. The connectivity between user locations and SBSs is modeled by parameters $\ell=\big(\ell_{ij}\in \{0,1\}: i\in\mathcal{I}, j\in\mathcal{J} \big)$, where $\ell_{ij}\!=\!1$ only if cache $j$ can be reached from location $i$. The MBS is within the range of all users in $\mathcal{I}$.

\textbf{File Requests}. The system evolves in slots, $t=1,2,\dots,T$. Users submit requests for obtaining files from a library $\mathcal{N}$ of $N$ files with unit size\footnote{For simplicity, we assume that files have unit size; but the results can be readily extended for the case the files are of size $s\neq 1$.}. We denote with $r_{t}^{n,i}\!\in\!\{0,1\}$ the event that a request for file $n\in\mathcal{N}$ has been submitted by a user at location $i\in\mathcal{I}$ during slot $t$. At each slot we  assume that there is exactly one request.\footnote{We can also consider batches of requests. If the batch has 1 request from each location, it is biased to equal request rate at each location. An unbiased batch contains an arbitrary number of requests from each location. Our guarantees hold for unbiased batches of arbitrary (but finite) length.} From a different perspective, this means that the policy is applied after every request, exactly as it happens with the standard LFU/LRU-type of reactive policies, see \cite{giovanidis-mLRU, leonardi-implicit} and references therein. Hence, the request process can be described by a sequence of vectors $\{r_t\}_{t=1}^{T}$ drawn from:
\begin{equation}
	\mathcal{R}=\Big\{r\in \{0,1\}^{N\times I} ~\Big |~ \sum_{n\in\mathcal{N}}\sum_{i\in\mathcal{I}} r^{n,i}=1\Big\}. \notag
\end{equation}

The instantaneous file popularity is expressed by the probability distribution $P(r_t)$ (with support $\mathcal{R}$), which is considered unknown and arbitrary. The same holds for the joint distribution $P(r_1,\dots,r_T)$ that describes the file popularity evolution within the time interval $T$. This general model captures all studied request sequences in the literature, including stationary (i.i.d. or otherwise), non-stationary, and adversarial models. The latter are the most demanding models one can employ as they include request sequences selected by an \emph{adversary} aiming to disrupt the system performance, e.g., consider Denial-of-Service attacks. If a policy achieves a certain performance under this model, it is guaranteed to meet this benchmark for all request models.
 
\textbf{Caching}. Each SBS $j$ can cache only $C_j$ files, with $C_j\!<\!N, \forall j\!\in\!\mathcal{J}$, while the MBS can store the entire library, i.e., $C_0\!=\!N$. One may also assume that the MBS has high-capacity direct access to the file server. Following the standard femtocaching model \cite{femtocaching}, we perform caching using the \emph{Maximum Distance Separable} (MDS) codes, where files are split into a fixed number of $F$ chunks, and each stored chunk is a pseudo-random linear combination of the original $F$ chunks. Using the properties of MDS codes, a user will be able to decode the file (with high probability) if it receives any $F$ coded chunks, a property that greatly facilitates cache collaboration and improves efficiency. 

The above model results in the following: the caching decision vector $y_t$ has $N\!\times\!J$ elements, and each element $y_t^{n,j}\in [0,1]$ denotes the amount of random coded chunks of file $n$ stored at cache $j$.\footnote{The fractional model is justified by the observation that large files are composed of thousands chunks, stored independently~\cite{maggi}.  Hence, rounding the fractional decisions  to the closest integer induces small errors.} Based on this, we introduce the set of eligible caching vectors: 
\[
\mathcal{Y}=\Big\{y\in [0,1]^{N\times J} ~\Big|~ \sum_{n\in\mathcal{N}}y^{n,j}\leq C_j, ~j\in J\Big\},
\]
which is convex. We can now define the online caching policy: 
\begin{definition}
A caching policy $\sigma$ is a (randomized) rule:
\begin{equation}
	\sigma: (r_1, r_2, \ldots, r_{t-1}, y_1, y_2, \ldots, y_{t-1})\longrightarrow y_{t}\in \mathcal{Y}\,. \nonumber
\end{equation}
which at each slot $t$ maps past observations $\{r_t\}_{t=1}^{t-1}$ and configurations $\{y_t\}_{t=1}^{t-1}$ to a new caching vector $y_t\in \mathcal{Y}$.
\end{definition}
\noindent Note that unlike previous \emph{strictly proactive} caching policies, we assume here that files can be cached dynamically in response to submitted requests.

\textbf{Routing}. Since each location $i\in\mathcal{I}$ is possibly connected to multiple caches, we introduce  \emph{routing variables} to determine the cache from which  the requested file will be fetched. Namely, let $z^{n,i,j}_t\!\in\![0,1]$ denote the portion of request $r_{t}^{n,i}$ that is served by cache $j$, and we define the respective routing vector $z_t$. There are two important remarks here. First, due to the coded caching model, the requests can be simultaneously routed from multiple caches. Second, the caching and routing decisions are coupled and constrained: \emph{(i)} a request cannot be routed from an unreachable cache; \emph{(ii)} we cannot route from a cache more data chunks than it has; and \emph{(iii)} each request must be fully routed, i.e., satisfied.

Based on the above, we define the set of eligible routing vectors conditioned on caching policy $y_t$ as:
\[
{\cal Z}(y_t)\!=\!\left\{z\!\in [0,1]^{N\!\times\! I\!\times \!J} \Bigg|
\begin{array}{c}
\sum_{j\in {\cal J}\cup\{0\}}z^{n,i,j}_t=r^{n,i}_t, \vspace{1.5mm}\\

z^{n,i,j}_t\!\leq\! \ell_{ij}y^{n,j}, ~\forall n\!\in\mathcal N,i,j\!\in \mathcal{J}\!\!
\end{array}
\right\}
\]
where the first constraint ensures that the entire request is routed, and the second constraint captures connectivity and caching limitations. We note that  routing from MBS (variable $z^{n,i,0}_t$) does not appear in the second constraint because the MBS stores the entire file library and can serve all users. This \emph{last-resort} routing option ensures that the set ${\cal Z}(y_t)$ is non-empty for all $y_t\in \mathcal{Y}$. As it will become clear in the next section, the optimal routing decisions can be easily devised for a given caching and request vector. This is an inherent property of uncapacitated bipartite caching networks, and also appears in prior works, e.g., see \cite{femtocaching}.   


\section{Problem Statement \& Formulation}\label{sec:formulation}
 
We begin this section by defining the caching objective and then proving that the online wireless edge caching operation can be modeled as a regret minimization problem.  

\begin{figure}[!t]
	\centering
	\includegraphics[width=3.5in]{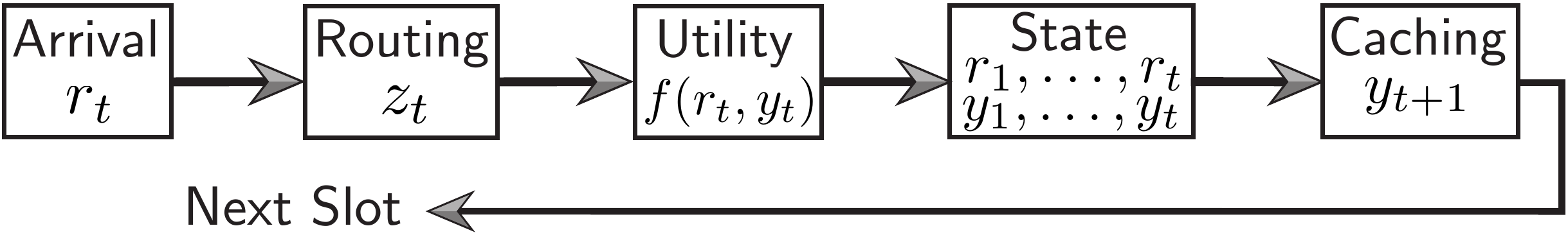} 
	\caption{\textbf{Online caching model}. When a request $r_t$ arrives, the file is routed optimally based on the current cache configuration i.e., $z_t=\mathcal{Z}(y_t)$. We accrue utility $f(r_t, y_t)$ and the caching decisions are updated using the state that includes the observed requests and caching decisions.}
	\label{fig:model}
\end{figure} 

\subsection{Cache Utility}

We consider a utility-cache model which is more general than cache-hit maximization \cite{paschos-jsac}. We introduce the weights $w^{n,i,j}$ to denote the utility when delivering a unit of file $n$ (i.e., a coded chunk) to location $i$ from cache $j$ instead of the MBS, and trivially set $w^{n,i,0}=0$. This detailed file-dependent utility model can be used to capture bandwidth economization from cache hits \cite{maggi}, QoS improvement from using caches in proximity \cite{femtocaching}, or any other cache-related benefit such as transmission energy savings due to proximity with the SBSs.\footnote{We can obtain the special case of hit ratio maximization from the above model if we set $w^{n,i,j}=w, \forall n,i,j$.} Our model allows these benefits to be different for each cache and user location due to, for example, the impact of wireless links; and we extend it in Sec. \ref{sec:extensions} to account for network dynamics such as link capacity variations. 

We can then define the network utility accrued in slot $t$ as: 
\begin{equation}\label{eq:biput}
	f_t(y_t)=\max_{z\in {\cal Z}(y_t)} \sum_{n\in\mathcal{N}}\sum_{i\in\mathcal{I}}\sum_{j\in\mathcal{J}} w^{n,i,j} r_t^{n,i}z^{n,i,j}_t,
\end{equation}
where index $t$ is used to remind us that $f_t$ is affected by the request $r_t$. It is easy to see that $f_t(\cdot)$ states that utility $w^{n,i,j}$ is accrued when a unit of request is successfully routed to cache $j$ where file $n$ is available. Note also that we have written $f_t(\cdot)$ only as function of caching, as for each $y_t$ we have already included in the utility definition the selection of the optimal routing $z_t$. As we will see next, this formulation facilitates the solution of the problem by simplifying the projection step.

\subsection{Problem Formulation}
 
Formulating the caching network operation as an OCO problem is non-trivial and requires certain conceptual innovations. For the discussion below please refer to Fig.~\ref{fig:model}. First, in order to model that the request sequence can follow any arbitrary and unknown probability distribution, we use the notion of an \emph{adversary} that selects $r_t$ in each slot $t$. In the worst case, this entity generates requests aiming to degrade the performance of the caching system. Going a step further, we model the adversary as selecting the utility function instead of the request. Namely, at each slot $t$, the adversary picks $f_t(y)$ from the family of  functions $\{f(r_t,y)\}$ by deciding the vector $r_t$. We emphasize that these functions are piece-wise linear. In the next subsection we will show that they are concave in the caching vector $y_t$, but not always differentiable. 

It is important to emphasize that we consider here the practical online setting where $y_t$ \emph{is decided after the request has arrived} and the caching utility has been calculated. This timing reflects naturally the operation of reactive caching policies, where first a generated request yields some utility (based on whether there was a cache hit or miss), and then the system reacts by updating the cached files. In other words, caching decisions are taken without knowing the future requests. The above steps allow us to reformulate the caching problem and place it squarely on the OCO framework \cite{hazan-book}.

Given the adversarial nature of our request model, the ability to extract useful conclusions depends crucially on the choice of the performance metric. Differently from the competitive ratio approach of \cite{Sleator85}, we introduce a new metric for designing our policies. Namely, we will compare how our policy fares against the best \emph{static} cache configuration designed in hindsight. This benchmark is a hypothetical policy that makes one-shot caching decisions having a priori knowledge of the entire request sequence. This metric is commonly used in machine learning \cite{hazan-book, Mert18} and is known as the worst-case static regret. In particular, we define the \emph{regret of policy $\sigma$} as:
\begin{align}\label{eq:regret}
	R_T(\sigma)&=\max_{P(r_1,\dots,r_T)}
	\mean{\sum_{t=1}^T f_t\big(y^*\big)-\sum_{t=1}^T f_t\big(y_t\big)},
\end{align}
where $T$ is the time horizon of reference. The maximization is over all possible adversary distributions and the expectation is taken w.r.t. the possibly randomized $r_t$ and $\{y_t, z_t(y_t)\}_{t=1}^T$. Essentially, this captures that the adversary can select any sequence of functions $\{f_t\}_{t=1}^T$ so as to deteriorate the effectiveness of our caching decisions\footnote{In defining the regret \cite{hazan-book}, the maximization is taken w.r.t. the sequence of functions, which for our problem is determined by the sequence of requests.}  $\{y_t\}_{t=1}^T$.

The best cache configuration is found by using the entire sample path of requests $\{r_t\}_{t=1}^T$ and solving:
\begin{equation}
y^*\!\in\!\arg\max_{y\in \Yc}\sum_{t=1}^T f_t(y). \notag
\end{equation}
Intuitively, measuring the performance of $\sigma$ w.r.t $y^*$ constrains the power of the adversary; for example a rapidly changing request pattern will impact $\sigma$ but also $y^*$. This comparison makes regret different from the standard competitive hit ratio\footnote{As explained in Sec. \ref{sec:related}, the competitive ratio metrics typically use a \emph{dynamic} benchmark that has full knowledge of requests and can select the exact optimal sequence of caching decisions, not simly a static configuration.} \cite{Sleator85}, and allow us to discern policies that learn high-utility caching configurations from those that fail to do so.

Our goal is to study how the regret scales with $T$. A policy $\sigma$ with sublinear regret produces average loss
\begin{equation}
\lim_{T\rightarrow\infty} R_T(\sigma)/T=0, \notag
\end{equation}
w.r.t. the optimal static cache configuration. This means that the two policies have the same average per-slot performance in the long-run, a property that is called \emph{no-regret} in OCO. In other words, $\sigma$ learns which file chunks to store and how to route requests, without having \emph{a priori} access to the file popularity. We can now formally define the online caching and routing problem as follows:
\begin{tcolorbox}[boxrule=1.1pt,arc=0.6em, title={Online Caching Problem (OCP)}] 
Given a file library $\mathcal{N}$; a set of user locations $\mathcal{I}$ and caches $\mathcal{J}$; a set of links connecting them $(\ell_{ij}:i\!\in\!\mathcal{I}, j\!\in\!\mathcal{J})$; and utilities $( w^{n,i,j}:n\!\in\!\mathcal{N}, i\!\in\!\mathcal{I}, j\!\in\!\mathcal{J} )$:\vspace{1.5mm}\\
Determine the policy $\sigma$ that selects at each slot  caching decisions $y_t$ that incur no regret over horizon $T$, i.e.,
$R_T(\sigma) = o(T)$, where $R_T(\sigma)$ is defined in \eqref{eq:regret}.
\end{tcolorbox}
We stress that while OCO typically focuses on time horizon $T$, in (OCP) the number and size of caches and, importantly, the library size, are large enough to induce high utility loss themselves. Hence, it is crucial to study how the regret is affected by these parameters as well.

\subsection{Problem Properties}
 
We prove that (OCP) is an OCO problem by establishing the concavity of $f_t(y)$ with respect to $y_t$. Note that we propose here a different formulation from the typical femtocaching model \cite{femtocaching} by including routing variables and the request arrival events. This re-formulation is imperative in order to fit our problem to the OCO framework, but also because otherwise (e.g., if were using \cite{femtocaching} ) we would need to make in each slot a computationally-challenging projection operation.

First, we simplify $f_t(y)$ by exploiting the fact that there is only one request at each slot. Let $\hat{n}, \hat{i}$ be the file and location where the request in $t$ arrives. Then $f_t(y_t)$ is zero except for $r_{t}^{\hat{n}, \hat{i}}$. Denoting with ${\cal J}_{\hat{i}}\subseteq \mathcal{J}$ the set of reachable SBS caches from $\hat{i}$, and simplifying the notation by setting $w^{\hat{n}, \hat{i}, j}=w^j$, $z^{\hat{n}, \hat{i}, j}=z^j$, and dropping subscript $t$, eq. \eqref{eq:biput} reduces to: 
\begin{align}
	f( y )\triangleq\,\,\,\,\,\,\max_{z\geq 0 }  \,\,\,&\sum_{j\in {\cal J}_{\hat{i}}} w^jz^j \label{eq:opt_zb1}\\
	\text{s.t. }& \sum_{j\in \mathcal{J}_{\hat{i}}} z^j\leq1 \label{eq:opt_zs}\\
	&  z^j\leq \left\{\begin{array}{ll}
		y^j, &  j\in {\cal J}_{\hat{i}}, \\
		0, & j\notin {\cal J}_{\hat{i}}.
	\end{array}\right. \label{eq:opt_zc}.
\end{align}

Although the standard femtocaching problem is known to be convex \cite{femtocaching}, our re-formulated utility function \eqref{eq:opt_zb1}-\eqref{eq:opt_zc} is different, and hence we need to show that $f(\cdot)$ remains convex.  
\begin{lemma}\label{lem:convx}
Function $f(y)$ is concave in its domain $\Yc$.
\end{lemma}
\begin{IEEEproof}
Consider two feasible caching vectors $y_1,y_2\in\Yc$. We will show that:
\[
f(\lambda y_1+(1-\lambda)y_2)\geq \lambda f(y_1)+(1-\lambda) f(y_2),~\forall \lambda \in [0,1].
\]
We begin by denoting with $z_1$ and $z_2$ the routing vectors that maximize \eqref{eq:opt_zb1} for vectors $y_1$, $y_2$ respectively. Immediately, it is $f(y_i)=\sum_j w^jz_i^j, ~i=1,2$. Next, consider a candidate vector
$y_3=\lambda y_1+(1-\lambda)y_2$, for some $\lambda\in [0,1]$. We first show that routing $z_3=\lambda z_1+(1-\lambda) z_2$ is feasible for $y_3$. By the feasibility of $z_1$, $z_2$, we have: 
\begin{align}
	\sum_j z_3^j\!=\!\sum_j (\lambda z_1^j\!+\!(1\!-\!\lambda) z_2^j)\! =\! \lambda \sum_j z_1^j+(1\!-\!\lambda)  \sum_j z_2^j\!\leq\!1, \nonumber
\end{align}
which proves that $z_3$ satisfies \eqref{eq:opt_zs}. Also, $\forall j\in \mathcal J$ it is:
\begin{align*}
	z_3^j&=\lambda z_1^j+(1-\lambda) z_2^j \leq \lambda y_1^j+(1-\lambda) y_2^j=y_3^j,
\end{align*}
thus $z_3$ satisfies also \eqref{eq:opt_zc} and $z_3\in \mathcal{Z}(y_3)$. It follows:
\begin{equation}
f(y_3)\triangleq  \max_{z\in \mathcal{Z}(y_3)}\sum_j w^j z^j\geq \sum_j w^j z^j_3. \nonumber
\end{equation}
Combining the above, we obtain:
\begin{align}
&f\big(\lambda y_1\!+(1\!-\lambda)y_2\big)=f(y_3)\geq \sum_j w^j z^j_3=\nonumber \\
&\lambda \sum_j w^j z^j_1+(1\!-\! \lambda) \sum_j w^j z^j_2\! =\!\lambda f(y_1)+(1\!-\!\lambda) f(y_2) \nonumber
\end{align}
which establishes the concavity of $f(y)$.
\end{IEEEproof}
Observe that the term $-\sum_{t=1}^Tf_t(y_t)$ of the regret definition is convex, and the $\max$ operator applied for all possible request arrivals preserves this convexity. This makes (OCP) an OCO problem, and this holds even when we consider general graphs and other convex functions $f_t(y_t)$.%

Finally, we can show with a simple example that $f_t(\cdot)$ does not belong to the class $\mathbf{C}^1$, i.e., it is not always differentiable. Consider a network with a single file $N\!=\!1$, and two caches with $C_1\!=\!C_2\!=\!1$, that serve one user with utility $w^1\!=\!w^2\!=\!1$. Assume that $y_t^{1,1}\!=\!y_t^{1,2}\!=\!0.5-\epsilon$ for some very small $\epsilon$. Notice that the partial derivatives $\partial f/\partial y_t^{1,1}=\partial f/\partial y_t^{1,2}=1$ (equal to $w$). But if we suppose a slight increase in caching variables such that $\epsilon$ term is removed, then the partial derivatives become zero. This is because extra caching of this file cannot improve the utility, which is already maximal. The same holds in many scenarios which make it impossible to guess when the objective changes in a non-smooth manner (having points of non - differentiability). Hence we will employ supergradients.

%% file: bipartite_R1.tex
\section{Bipartite Supergradient Caching Algorithm}\label{sec:bipartite}

Our solution employs an efficient and lightweight gradient-based algorithm for the caching decisions, which incorporates the optimal routing as a subroutine. We start from the latter.

\subsection{Optimal Routing}

Recall that file routing is naturally decided after a request is submitted, at which time the caching $y_t$ has been determined, Fig.~\ref{fig:model}. Thus, in order to decide $z_t$ we will assume $r_t$ and $y_t$ are given. The goal of routing is to determine which chunks of the file are fetched from each cache. 

Specifically, let us fix a request for file $\hat{n}$ submitted to location $\hat{i}$. Using the notation $\mathcal{J}_{\hat i}, w^j$ defined above, and  
letting $\hat{z}\triangleq z^{\hat{n},\hat{i},j}$ be the optimal routing variables related to these caches, file, and user location, we may recover an optimal routing vector as one that maximizes the utility: 
\begin{align}
	\hat{z}\in \,\,\,\,\,\,{\arg\max}_{z\geq 0, \eqref{eq:opt_zs},\eqref{eq:opt_zc} }  \,\,\,&\sum_{j\in {\cal J}_{\hat i}} w^jz^j. \label{eq:opt_routing}
\end{align} 
Ultimately, the routing at $t$ is set:
\[
z_t^{n,i,j} = \left\{\begin{array}{ll}
\hat{z}^{j} & \text{if } n=\hat{n},~ i=\hat{i}, \\
0 & \text{otherwise}.
\end{array}\right.
\]
Problem \eqref{eq:opt_routing} is a Linear Program (LP) of a dimension at most $\text{deg}$, where $\text{deg}\!=\!\max_{i\in\mathcal{I}} |\mathcal{J}_i|$, and $|\mathcal{J}_i|$ is the number of caches reachable from location $i$. This LP is computationally-efficient and can be solved by the interior point or the simplex method \cite{bertsimas1997introduction}. Interestingly, however, due to its structure a solution can be found by inspection as follows. First, we order the reachable caches in decreasing utility, i.e., let $\phi(\cdot)$ be a permutation such that $w^{\phi(1)}\geq w^{\phi(2)} \geq \dots \geq w^{\phi(|\mathcal{J}_{\hat{i}}|)}$. We set $z^{\phi(1)}=\min\{1,y^{\hat{n},\hat{i},\phi(1)}\}$ for the first element, and then iteratively for each round $k$, we set:
\begin{equation}
	z^{\phi(k)}=\min\{y^{\hat{n},\hat{i},\phi(k)}, 1-\sum_{i=1}^{k-1} z^{\phi(i)}\}, \nonumber
\end{equation}
until all reachable caches are visited, or we obtain $\sum_{i=1}^k z^{\phi(k)}\!=\!1$ for some $k$; where in the latter case the rest of the caches have $z^j\!=\!0$. Both approaches, i.e., solving directly the LP or using this iterative process, may be helpful in practice. By explicitly solving the LP we also obtain the value of the dual variables, which, as we will see, help us to compute the supergradient.

\subsection{Optimal Caching - BSCA Algorithm}
The general idea is to gradually update caching decisions along the direction of the gradient. However, since $f(y)$ is not differentiable everywhere we need to find a supergradient direction at each slot. We describe next how this can be achieved. Consider the partial Lagrangian of \eqref{eq:opt_zb1}: 
\begin{equation}
L(y,z,\alpha,\beta)\!=\!\sum_{j\in\mathcal{J}_{\hat{i}}}\!w^jz^j\!+\!\alpha\big(1-\!\sum_{j\in\mathcal{J}_{\hat{i}}} \!z^j\!\big)+\sum_{j\in\mathcal{J}_{\hat{i}} }\!\beta^j(y^j\!-\!z^j) \label{eq:lagrangian}
\end{equation}
where $w^j\triangleq w^{\hat{n}, \hat{i},j }$, and define the auxiliary function:
\begin{equation}
\Lambda (y,\beta)=L(y,z^*,\alpha^*,\beta)\triangleq \min_{\alpha\geq 0}\max_{z\geq 0}	L(y,z,\alpha,\beta). \label{eq:lambda}
\end{equation}
From the strong duality property of linear programming \cite{bertsimas1997introduction}, we may exchange $\min$ and $\max$ in the Lagrangian, and  obtain:
\begin{equation}\label{eq:strdual}
f(y)=\min_{\beta \geq 0}\Lambda (y,\beta).
\end{equation}

\noindent We prove next the following lemma for the supergradients.
\begin{lemma}[Supergradient]\label{lem:subgradient}
Let $\beta^*(y)\!\triangleq\! \arg\min_{\beta\geq 0}\!\Lambda (y,\beta)$ be the vector of optimal multipliers corresponding to  \eqref{eq:opt_zc}. Define
\begin{align}\label{eq:supergrad}
g^{n,i,j} = \left\{\begin{array}{ll}
\beta^{j,*}(y) & \text{if } n=\hat{n},~ i=\hat{i}, ~j\in \mathcal{J}_{\hat{i}} \\
0 & \text{otherwise}.
\end{array}\right.
\end{align}
The vector $g\!\in\!\mathbb R^{N\times I \times J}$ is a supergradient of $f$ at $y$, i.e., it holds $f(y)\!\geq\! f(y')\!-g^\top(y'-y),~\forall y'\!\in\! \Yc$. 
\end{lemma}
\begin{IEEEproof}
First note that we can write: $f(y)\stackrel{\eqref{eq:strdual}}{=}$
\begin{align*}
	&\min_{\beta \geq 0}\Lambda (y,\beta)\triangleq \Lambda \big(y,\beta^*(y)\big) \stackrel{(a)}{=}\Lambda \big(y',\beta^*(y)\big)-\beta^*(y)^\top(y'-y).
\end{align*}
Where $(a)$ holds since it is:
\begin{align}  
L(y',z,\alpha,\beta)\!=\!\sum_{j\in\mathcal{J}_{\hat{i}}}\!w^jz^j\!+\!\alpha\big(1-\!\sum_{j\in\mathcal{J}_{\hat{i}}} \!z^j\!\big)+\sum_{j\in\mathcal{J}_{\hat{i}} }\!\beta^j(y'^{,j}\!-\!z^j)\nonumber
\end{align}
and by applying \eqref{eq:lambda}, where the optimization is independent of variables $y$ (or $y'$), we obtain $\Lambda(y',\beta)=L(y', z^*, \alpha^*, \beta)$, with $\alpha^*$ and $z^*$ being the same as those appearing in $\Lambda(y,\beta)=L(y, z^*, \alpha^*, \beta)$ (since their calculation is independent of $y$). Hence, we can subtract the two expressions (observe the linear structure of \eqref{eq:lagrangian}), plug in a certain vector $\beta$ and obtain:
\begin{equation}
	\Lambda\big(y,\beta^*(y)\big)-\Lambda\big(y', \beta^*(y)\big)=-\beta^*(y)^\top(y'-y).
\end{equation}
where $\beta^*(y)\!=\!\arg\min \Lambda(y,\beta)$. Note also that it holds $\Lambda (y',\beta^*(y))>\Lambda (y',\beta^*(y'))$ by definition of $\beta^*$, hence:
\begin{align}
f(y)&=\Lambda \big(y',\beta^*(y)\big)-\beta^*(y)^\top(y'-y)\nonumber \\
 &\geq \Lambda\big(y', \beta^*(y')\big)-\beta^*(y)^\top(y'-y) \nonumber \\
&=f(y')-\beta^*(y)^\top(y'-y), \nonumber
\end{align}
which concludes our proof.
\end{IEEEproof}

Intuitively, the dual variable $\beta^{j,*}(y)$ (element of vector $\beta^*(y)$) is positive only if the respective constraint \eqref{eq:opt_zc} is tight which ensures that increasing the allocation $y^{\hat{n},\hat{i},j}$ will induce a benefit in case of a request with $r^{\hat{n},\hat{i}}\!=\!1$ occurs in future. The actual value of $\beta^{j,*}(y)$ is proportional to this benefit. The reason the algorithm emphasizes this request, is that in the online gradient-type of algorithms the last function (in this case a linear function with parameters the last request) serves as a corrective step in the ``prediction'' of future. Having this method for calculating a supergradient direction, we can extend the seminal online gradient ascent algorithm \cite{zinkevich2003online}, to design an online caching policy for (OCP). In detail: 
\begin{definition}[BSCA]
The {Bipartite Subgradient Caching Algorithm} adjusts the caching decisions with a supergradient: 
\[
y_{t+1}=\Pi_{\mathcal{Y}}\left(y_t+\eta_t g_t\right),
\]
where $\eta_t$ is the stepsize, $g_t$ can be taken as  in Lemma~\ref{lem:subgradient}, and
\begin{equation}
\Pi_{\mathcal{Y}}\left(q\right)\triangleq \argmin_{y\in\Yc}\|q-y \|, \label{eq-projection1}
\end{equation}
is the Euclidean projection of the argument vector $q$ onto $\mathcal{Y}$.
\end{definition}

Algorithm \ref{alg1} explains how BSCA can be incorporated into the network operation for devising the caching and routing decisions in an online fashion. The algorithm requires as input only the network parameters $\ell_{ij}, C_j, \mathcal{N}, w^{n,i,j}$. The stepsize $\eta_t$ is computed using the set diameter $\Delta_{\mathcal{Y}}$, the upper bound on the supergradient  $K$, and the time horizon $T$. The former two depend on the network parameters as well. Specifically, define first the diameter $\Delta_{\mathcal{S} }$ of  set $\mathcal{S}$ to be the largest Euclidean distance between any two elements of this set. In order to calculate this quantity for $\mathcal{Y}$, we select vectors $y_1, y_2\in \Yc$ which cache exactly $C_j$ different files at each cache $j\!\in\! \mathcal{J}$, and hence:
\begin{equation}\label{eq:diam}
\Delta_{\mathcal{Y}}=\sqrt{\sum_{n\in\mathcal N}\sum_{j\in\mathcal J}(y_1^{n,i,j}-y_2^{n,i,j})^2}=\sqrt{\sum_{j\in \mathcal{J}}2C_j}\leq \sqrt{2CJ}, \notag
\end{equation}
where $C\!\triangleq\!\max_{j}C_j$. Also, we denote with $K$ the upper bound on the norm of the supergradient vector. By construction this vector is non-zero only at the reachable caches, and only for the specific file. Further, its smallest value is zero by the non-negativity of Lagrangian multipliers, and its  largest is no more than the \textbf{maximum utility}, denoted with $w^{(1)}$. Thus, using $\text{deg}\!=\!\max_{i\in\mathcal I} |\mathcal{J}_i|$ we can bound the supergradient norm: 
\begin{equation}\label{eq:lipschitz}
\|g\|= \sqrt{\sum_{j\in \mathcal{J}_{\hat{i}}} (w^{(1)})^2} \leq w^{(1)}\sqrt{\text{deg}}\triangleq K.
\end{equation}

\begin{algorithm}[t]
	\nl \textbf{Input}: $\{\ell_{ij}\}_{(i,j)}$; $\{C_j\}_j$; $\mathcal{N}$; $\{w^{n,i,j}\}_{(n,i,j)}$; $\eta_t=\Delta_{\Yc}/K\sqrt{T}$.\\%
	\nl \textbf{Output}: $y_t$, $\forall t$.\\%
	\nl \textbf{Initialize}: $\hat{n}, \hat{i}$, $y_1$ arbitrarily.\\%
	\nl \For{ $t=1,2,\ldots $  }{
		\nl Observe  request $r_{t}$ and set ${\hat{n}, \hat{i}}$ for which  $r_{t}^{\hat{n},\hat{i}}\!=\!1$\,;\\
		\nl Find the  routing $z_t$ solving \eqref{eq:opt_zb1}-\eqref{eq:opt_zc}; $\%$ \emph{decides routing}		\\
		\nl Calculate the accrued utility $f_t(r_t,y_t)$\,;\\
		\nl Calculate the supergradient $g_t$ for $\hat{n}, \hat{i}$ using (\ref{eq:supergrad}); \\%
		\nl Update the vector $q_{t+1}=y_{t}+\eta_tg_t$\,; \\%
		\nl Project: $y_{t+1}=\Pi_{\mathcal{Y}}\left(q_{t+1}\right)$; \,\,$\%$ \emph{decides caching}\\%
	}
	\caption{{\small \!Bipartite Supergradient Caching Algorithm }}	\label{alg1}
\end{algorithm} 

The algorithm proceeds as follows. At each slot $t$, the system receives a request $r_t$ and sets $\hat{i}, \hat{n}$ for the requester and file (line 5). Given the cached files, the system finds the optimal routing $z_t$ for serving $r_t$ (line 6), e.g. by solving an LP with at most $\text{deg}$ variables and finding the dual variables. This yields utility $f_t(r_t,y_t)$ (line 7). The supergradient is calculated (line 8) and is used to update the cache configuration (line 9). Finally, the decisions are projected to the feasible set so as to satisfy the cache capacities (line 10).

It is interesting to note the following. Since the supergradient computation in line 8 and the optimal routing, explained in the previous subsection, require the solution of the same LP, it is possible to combine these as follows. When the optimal routing is found, the dual variables are stored and used for the direct computation of the supergradient in the next iteration of BSCA. Note that, given the cache update rule, the algorithm state needs to include only $y_t$, and therefore its memory requirements are very small.

\subsection{Performance of BSCA}
Following the rationale of the analysis in \cite{zinkevich2003online}, we show that our policy achieves no regret and we analyze how the various system parameters affect the regret expression.
\begin{tcolorbox}[boxrule=1.1pt,arc=0.6em, left=0.25mm, right=0.25mm] 
\begin{theorem}\label{th:3}
The regret of \textup{BSCA} satisfies:
\[
R_T(\textup{BSCA})\leq 
w^{(1)}\sqrt{2\textup{deg} J CT},\,\,\,\,\,\,\,\text{where}
\]
$C\!=\!\max_j C_j$,\, \text{deg}$=\max_{i\in\mathcal I} |\mathcal{J}_{i}|$,\, $w^{(1)}\!=\!\max_{n,i,j}w^{n,i,j}$
\end{theorem}
\end{tcolorbox}
\vspace{1mm} 
\begin{IEEEproof}
Using the non-expansiveness property of the Euclidean projection \cite{Ber99book}, we can bound the distance of each new value $y_{t+1}$ from the hindsight policy $y^*$, as follows:
\begin{align}
&\|\Pi_{\mathcal{Y}}\left(y_t\!+\!\eta_t  g_t\right)\!-\!y^*\|^2\!\leq\! \|y_{t}\!+\!\eta_tg_t\!-\!y^*\|^2\!=\! \nonumber \\ &\|y_t\!-\!y^*\|^2\!+\!2\eta_t{g_t}^\top(y_t\!-\!y^*)\!+\!\eta_t^2\|g_t\|^2,\label{eq:ineqsub1}
\end{align}
where we expanded the norm. If we fix the step size $\eta_t=\eta$ and sum telescopically over all slots until $T$, we obtain:
\begin{equation*}
\|y_{T+1}-y^*\|^2\!\!\leq\! \|y_1-y^*\|^2+2\eta\sum_{t=1}^T{ g_t}^\top(y_t-y^*)+\eta^2\sum_{t=1}^T\!\|g_t\|^2.
\end{equation*}
Since $\|y_{T+1}-y^*\|^2\geq 0$, rearranging the terms and using that $\|y_1-y^*\|\leq \Delta_{\Yc}$ and $\|g_t\|\leq K$ we obtain:
\begin{equation}\label{eq:teleonl}
\sum_{t=1}^T{g_t}^\top(y^*-y_t)\leq \frac{\Delta_{\Yc}^2}{2\eta} +\frac{\eta TK^2}2.
\end{equation}
	
Since our utility function is concave, it holds:
\begin{equation}	
f_t(y)\!\leq\! f_t(y_t)\!+\! {g_t}^\top(y-y_t), \,\,\,\forall y\in\mathcal{Y}
\end{equation} 
for every $f_t$, and therefore also for the function that maximizes the regret; thus, we can remove the $\max$ operator from \eqref{eq:regret} and rewrite it as:
\begin{align*} 
R_T(BSCA)&=\sum_{t=1}^T(f_t(y^*)-f_t(y_t)) \notag \\
&\leq \sum_{t=1}^T{g_t}^\top(y^*-y_t)\stackrel{\eqref{eq:teleonl}}{\leq} \frac{\Delta_{\Yc}^2}{2\eta} +\frac{\eta TK^2}2.
\end{align*}
We can minimize the regret bound by optimizing the step size. Using the first-order condition w.r.t. $\eta$ for the RHS of the above expression, we obtain $\eta^*=\Delta_{\Yc}/K\sqrt{T}$ which  yields:
\begin{align}
	R_T(BSCA)\leq  \Delta_{\mathcal{Y}}K\sqrt{T}. \label{eq:conv_reg}
\end{align}
Recall also that $\Delta_{\mathcal{Y}}\leq \sqrt{2CJ}$, and that $K= \omega^{(1)}\sqrt{\text{deg} }$, due to \eqref{eq:diam}-\eqref{eq:lipschitz}. Hence, the theorem follows.
\end{IEEEproof}

Theorem 1 shows that the regret of BSCA scales as $O(\sqrt{T})$ and therefore BSCA solves (OCP). The regret expression captures how fast the algorithm \emph{learns} the right caching configuration, and therefore the detailed constants we obtain in the theorem are of great importance. For example, we see that the bound is independent of the file library size $N$. This is very crucial in caching problems where the $N$ drives the problem's dimension. Another interesting observation is that the learning rate of the algorithm might become slow (i.e., resembling regret behavior of $\sim\!\!O(T)$) when $C$ is comparable to $T$. This is in line with empirical observations suggesting that in order to extract safe conclusions about the performance of a policy, one should simulate datasets with size $T>>C$. 

We stress also that Theorem 1 does not imply that BSCA outperforms all other possible policies; for example, if the requests have a particular structure, e.g., are highly correlated, then another policy might perform better. However, policies that exploit the structure of requests tend to perform poorly when the request model assumptions do not hold. We present such examples in Sec. \ref{sec:performance}

Finally, note that calculating $\eta^*\!=\!\Delta_{\Yc}/K\sqrt{T}$ requires to know $T$, but this can be relaxed by using the standard \emph{doubling trick} \cite{hazan-book}. Alternatively, we can employ a diminishing step. Namely, if we sum telescopically \eqref{eq:ineqsub1} for $T$ slots, we obtain:  
\begin{equation}\label{eq:2nd-ubound}
	R_{T}^v(BSCA)\leq  \frac{\Delta_{\mathcal{Y}}^2}{2\eta_{T}}+ \frac{ K^2 \sum_{t=1}^T\eta_t }{2},
\end{equation}
and if we set $\eta_t=1/\sqrt{t}$, then the two terms in \eqref{eq:2nd-ubound} yield factors of order $O(\sqrt{T})$, hence:
\begin{equation}\label{eq:22nd-ubound}
R_{T}^v(BSCA)\leq  \frac{\Delta_{\mathcal{Y}}^2 \sqrt{T}}{2}+ \Big(\sqrt{T}-\frac{1}{2} \Big)K^2. 
\end{equation}
Comparing the two sublinear regret expressions, \eqref{eq:conv_reg} and \eqref{eq:22nd-ubound}, we see that their exact relationship depends on the relative values of parameters $K$ and $\Delta_{\mathcal{Y}}$.

%% file: projection_R1.tex
\section{Cache Projection Algorithm}\label{sec:projection}

BSCA involves a projection (line 9) which might affect significantly its complexity and runtime. We develop here a tailored algorithm that resolves this issue. 

The Euclidean projection defined in \eqref{eq-projection1} can be written as the equivalent quadratic program:
\begin{align}
\Pi_{\mathcal{Y}}\left(q\right)\triangleq\,\,\,\, \argmin_{y} & \sum_{j\in\mathcal{J}}\sum_{n\in\mathcal{N}} (y^{n,j}-q^{n,j})^2 \label{eq:projection}\\
\text{s.t.    }& \sum_{n\in\mathcal{N}} y^{n,j} \leq C_j,\,\,\,\,\,\,\forall j\in\mathcal{J} \nonumber \\
&\,0\leq y^{n,j}\leq 1,\qquad\forall n\in\mathcal{N}, j\in\mathcal{J},\notag 
\end{align} 
that might be computationally very expensive in some cases; see \cite{maculan-projection, wang2015projection} and references therein. Our problem has certain properties that facilitate this operation. First, the projection can be performed independently for each cache, namely we project $q$ on the intersection of a simplex-type constraint $\sum_n y^{n,j}\leq C_j$ and a $N$-dimensional box $y^{j}\in [0,1]^N$ (\emph{capped simplex}). Second, $y_{t+1}^j$ and $q_{t+1}^j$ differ only in one element. Exploiting these properties we design an algorithm for (\ref{eq:projection}) with complexity $O(JN)$, that uses the Karush-Kuhn-Tucker (KKT) conditions \cite{Ber99book} to navigate fast the solution space. 




We first introduce the Lagrangian:
\begin{align}
L(y, \rho, \mu, \kappa )&\!=\!\sum_{n\in \mathcal{N}}\sum_{j\in\mathcal{J}}(q^{n,j}\!-\!y^{n,j})^2+\sum_{j\in\mathcal{J}}\rho_j\Big(\sum_{n\in\mathcal{N}}y^{n,j}\!-\!C_j\Big) \nonumber \\ 
&+\sum_{n\in\mathcal{N} }\sum_{ j\in\mathcal{J} }\mu_{n,j}(y^{n,j}-1) - \sum_{n \in\mathcal{N} }\sum_{j\in\mathcal{J}}\kappa_{n,j}y^{n,j}\nonumber
\end{align}
where $\rho_j,\mu_{n,j},\kappa_{n,j}$, $\forall n,j$, are the non-negative Lagrange multipliers introduced when relaxing the constraints above. The KKT conditions of \eqref{eq:projection} at the optimal point, are:
\begin{align}
	& \frac{ \partial L(\cdot) }{\partial y^{n,j} }\!=\!-2(q^{n,j}-y^{n,j})+\rho_j\!+\!\mu_{n,j}\!-\!\kappa_{n,j}\!=\!0,\,\,\forall n,j, \label{eq:KKT-1a}
\end{align}
\begin{align}	
	& \rho_j\big( \sum_{n\in\mathcal{N}}y^{n,j}-C_j \big)=0,\,\,\,\,\,\forall j, \label{eq:KKT-2a} \\
	& \mu_{n,j}(y^{n,j}-1)=0,\,\,\,\, \kappa_{n,j}y^{n,j}=0\,\,\,\,\forall n,j, \label{eq:KKT-3a} \\
	& \rho_j,\,\,\mu_{n,j},\,\,\kappa_{n,j}\geq 0,\,\,\,\,\forall n, j, \label{eq:KKT-4a} 
\end{align}
where we have omitted the primal constraints of \eqref{eq:projection} for brevity. In order to solve the projection problem we will use a simple algorithm that tests, in a systematic fashion, combinations of the complementary slackness conditions \eqref{eq:KKT-2a}-\eqref{eq:KKT-3a} until it finds a solution that is primal and dual feasible. An important observation is the following: since $C\!<\!N$, the simplex constraint will be tight at the optimal point (the cache is filled) and hence we only need to check cases for \eqref{eq:KKT-3a}.

First, note that the caching decisions $y^j=(y^{n,j}, n\in\mathcal{N})$ for each cache $j$ are partitioned at the optimal point into three sets defined as follows:
\begin{align}
&\Mc_{1}^j=\{n\in\mathcal{N}: y^{n,j}=1 \},\,\,\Mc_{3}^j=\{n\in\mathcal{N}: y^{n,j}=0 \},\notag  \nonumber \\ &\Mc_{2}^j=\{n\in\mathcal{N}: y^{n,j}=q^{n,j}-\rho_j/2\},\nonumber 
\end{align}
where $\Mc_{1}^j$ contains the files that will be stored in their entirety, $\Mc_{2}^j$ the partially cached files ($\kappa_{n,j}=\mu_{n,j}=0$) in cache $j$, and $\Mc_{3}^j$ the evicted files. Due to full utilization of cache capacity, it holds for each cache $j\in\mathcal{J}$:
\begin{align}
&\sum_{n\in\mathcal{N}} y^{n,j}=C_j=|\Mc_{1}^j|+\sum_{n\in \Mc_{2}^j} q^{n,j} - \rho_j|\Mc_{2}^j|/2\,\, \Rightarrow \nonumber \\
&\rho_j=2\big(|\Mc_{1}^j|-C_j+\sum_{n\in \Mc_{2}^j} q^{n,j}\big)/|\Mc_{2}^j|. \notag 
\end{align}

\begin{algorithm}[t]
	\nl \textbf{Input}: $\{C_j\}_j$; $q^j,\forall j\in\mathcal{J}$; \,\,\,\,\,\textbf{Output}: $y = \Pi_{\mathcal{Y}}\left(q\right)$;\\
	\nl \textbf{Initialize}: $\Mc_{1}^j\leftarrow \emptyset,\Mc_{2}^j\leftarrow \mathcal{N},\Mc_{3}^j \leftarrow \emptyset$, $\forall j\in\mathcal{J}$.\\%
	\nl \Repeat{ $\mathcal{S}_j=\emptyset,\,\forall j\in\mathcal{J}$  }{
		\For{$j=1,2,\ldots,J$}{
			\nl $\rho_j\leftarrow 2({|\Mc_{1}^j|-C+\sum_{n\in \Mc_{2}^j} q^{n,j}})/{|\Mc_{2}^j|}$;\\
			\nl $y^{n,j}\leftarrow\left\{\begin{array}{ll}
			1 &  n\in \Mc_{1}^j, \\
			q^{n,j}-\rho_j/2 & n\in \Mc_{2}^j, \\
			0 & n\in \Mc_{3}^j \end{array}\right.$;\\
			\nl $\mathcal{S}_j \leftarrow \left\{n\in\mathcal{N}: y^{n,j}<0\right\}$ ; \\%
			\nl $\Mc_{2}^j \leftarrow \Mc_{2}^j\setminus \mathcal{S}_j$, $\Mc_{3}^j \leftarrow \Mc_{3}^j\cup \mathcal{S}_j$ ; \\%
			\nl \If{$y^{1,j} >1$}{$\Mc_{1}^j\leftarrow \{1\},\,\,\Mc_{2}^j\leftarrow \{2,\dots,N\},\,\,\Mc_{3}^j \leftarrow \emptyset$;\\
				GoTo line 5;} }
	}
	\caption{Fast Cache Projection}  \label{alg2}
\end{algorithm}\DecMargin{1em}

In order to solve the projection problem it suffices to determine for each cache a partition of files into sets $\Mc_{1}^j,\Mc_{2}^j,\Mc_{3}^j$. Note that we can check in linear time if a candidate partition satisfies all KKT conditions (and only the optimal one will). Additionally, one can show that the ordering of files in $q$ is preserved at optimal $y$, hence prior approaches, e.g., \cite{wang2015projection} that search exhaustively over all possible ordered partitions will need $O(N^2)$ steps. Here, however, we expedite the solution by exploiting the property that all elements of $q$ satisfy $q^{n,j}\leq 1$ except at most one (hence also $|\Mc_{1}^j|\in \{0,1\}$ for every cache $j$). This allows us to reduce the runtime to  $O(N\log N)$ steps for each cache.
Furthermore, our algorithm can also operate without sorting the files, and therefore the runtime for one cache is $O(N)$, and the overall runtime is $O(JN)$.

The details are presented in Algorithm \ref{alg2}. The initial partition places all files to the set of partially cached (line 2). For the given partition, we compute the Lagrange multiplier $\rho_j$ (line 4), and calculate a tentative caching allocation (line 5). The indices of all files whose tentative allocation is negative are stored in a set $S_j$ (line 6), removed from the middle set and added to the set of files to be evicted (line 7). If there exists a file with allocation more than 1, it is placed at the set of fully cached, and the procedure is repeated. We exploit the structure of our problem: since in the previous slot all files had allocation at most 1, it follows that adding the supergradient element and taking into account the multiplier $\rho_j$, the new allocation of all files (but the one in the supergradient) will be strictly smaller than 1. Therefore, $\mathcal{M}_1^j$ can either have one file or none, and we search between these two possibilities (line 8). The set operation we perform in line 7 is proven in \cite{wang2013projection} to be monotonous, and therefore we will at most search all possibilities, resulting in worst-case runtime $O(JN)$ that matches previously known results \cite{maculan-projection}. Finally, we observed in simulations that each loop was visited at most two times (instead of $N$), resulting in an extremely fast projection.

%% file: gradient_descent_R1.tex
\section{The Single Cache Case}\label{sec:gradient}

The problem is simplified for a single cache as there are no routing decisions. Nevertheless, even for this basic version, we lack a policy that can achieve no-regret caching performance for any request sequence. BSCA not only fills this gap, but in fact it achieves the best learning rate than \emph{any} possible policy (based on OCO or not) can achieve.\footnote{As it will become clear in the simulations, BSCA ensures no regret for any request sequence, and in the case of a single cache we prove that its learning rate is the best possible. However, this does not mean that there are no policies which can achieve better performance for specific request patterns.}

\subsection{BSCA for One Cache}
We denote with $C$ the size of our single cache and with $r_{t}^n$ the request arriving at slot $t$, where now we do not consider different user locations as all requests are served by the same cache. The cache utility can be written:
\begin{equation}
	f_t(y_t)=\sum_{n\in\mathcal{N}}y_{t}^nw^nr_{t}^n,
\end{equation}
which states that a request for file $n$ yields utility proportional to a file-specific parameter $w^n$ per unit of its cached fraction $y_{t}^n$. There are no routing variables in this case. Also, the gradient $g_t\triangleq\nabla f_t$ at $y_t$ exists, and it is the $N$-dimensional vector with coordinates:
\[
\frac{\partial f_t}{\partial y_t^n}=w^n r^n_t,~n=1,\dots,N.
\]
This simplifies the implementation of BSCA as we can directly calculate the gradients in each slot $t$ and update the cached files using $y_{t+1}=\Pi_{\mathcal{Y}}( y_t+\eta_t g_t)$.

The regret of BSCA for the one cache (henceforth called BSCA-1) stems from Theorem 1. Namely, setting $J\!=\!1$ and $\text{deg}\!=\!1$, we derive the following Corollary.
\begin{tcolorbox}[boxrule=0.7pt,arc=0.6em, colback=gray!3] 
\begin{corollary}[Regret of BSCA-1]\label{th:2}
Fix step size $\eta_t=\Delta_{\mathcal{Y}_1}/K_1\sqrt{T}$, the regret of BSCA for 1 cache satisfies:
\[
R_T(\textup{BSCA-1})\leq \Delta_{\mathcal{Y}_1}K_1{\sqrt{T}} \leq {w^{(1)}\sqrt{2CT}}
\]
\end{corollary}
\end{tcolorbox}
\noindent Where $K_1\!=\!\max_n\{w_n\}\!\triangleq\! w^{(1)}$ upper bounds $\nabla f_t,\,\forall t$; and the eligible caching vectors belongs to the convex set:
\[
\mathcal{Y}_1=\Big\{y\in [0,1]^{N} ~\Big|~ \sum_{n\in\mathcal{N}}y^{n}\leq C \Big\},
\]
which has dimension:
\[
\Delta_{\mathcal{Y}_1}=\left\{
\begin{array}{ll}
\sqrt{2C} & \text{if } 0<C\leq N/2,\\
\sqrt{2(N-C)} & \text{if } N/2<C\leq N. 
\end{array}
\right.
\]
It is easy to calculate $\Delta_{\mathcal{Y}_1}$ using two vectors $y_1, y_2\in\mathcal{Y}_1$ that cache entirely different files.

\subsection{Regret Lower Bound}\label{sec:regret-bound}

We now derive a regret lower bound which is a powerful theoretical result that provides the fundamental limits of how fast any algorithm can learn to cache, much like the information-theoretic upper bound of the channel capacity. Regret lower bounds in OCO have been previously derived for, e.g., $N$-dimensional unit ball centered at the origin in \cite{Abernethy08}, and  $N$-dimensional hypercube in \cite{hazan-book}. In our case, however, the above results are not tight since  $\mathcal{Y}_1$ is a capped simplex, i.e., the intersection of a box and a simplex inequality. Therefore, we need the following new regret lower bound tailored to the online caching problem.

\begin{tcolorbox}[boxrule=1.1pt,arc=0.6em, left=1mm, right=1mm] 
	\begin{theorem}[Regret Lower Bound]\label{th:lowerBoundRegret}
		The regret of any  caching policy $\sigma$   satisfies:
		\[
		R_T(\sigma) > \sum_{i=1}^C\mean{Z_{(i)}}\sqrt{T},\quad \text{ as } T\to\infty, \vspace{-0.051in}
		\]
		where $Z_{(i)}$ is the $i$-th max element of a Gaussian random vector with zero mean and covariance matrix $\Sigmam(w)$ in \eqref{eq:covarianceMatrix}.  
		
		Furthermore, assume $C<N/2$ and define $\phi$ any permutation of $\mathcal{N}$ and $\Phi$ the set of all such permutations, then:
		\[
		R_T(\sigma)>\frac{\max_{\phi\in\Phi}\sum_{k=1}^C\sqrt{w^{\phi(2(k-1) +1)}+ w^{\phi(2k)}}}{\sqrt{2\pi\sum_{n=1}^N1/w^n}}\sqrt{T}	
		\] 
	\end{theorem}
\end{tcolorbox}

In the special case that we wish to maximize the hit rate, where it is $w^n\!=\!w, \forall n\in\mathcal N$, the above bound simplifies to:
\begin{tcolorbox}[boxrule=0.7pt,arc=0.6em, colback=gray!3] 
	\begin{corollary}\label{cor:lbound1}
		Fix $\gamma\triangleq C/N$,  $w^n=w,~\forall n$, and $C<N/2$. Then, the regret of any caching policy $\sigma$  satisfies: 
		\[
		R_T(\sigma) > w\sqrt{\frac{\gamma}{\pi}}\sqrt{CT}, \quad \text{ as } T\to\infty.
		\]
	\end{corollary}
\end{tcolorbox}
\noindent This bound is tighter than the classical result $\Omega\big(\sqrt{T\log N}\big)$ \cite{hazan-book, Abernethy08}, which is attributed to the difference of  sets $\mathcal{R},\mathcal{Y}_1$.

\begin{IEEEproof} [Proof of Theorem \ref{th:lowerBoundRegret}]
To find a lower bound, we will analyze a specific adversary $r_t$. In particular, we will consider an i.i.d.  $r_t$ such that file $n$ is requested with probability: 
\[
\prob{r_t=\ev_n}=\frac{1/w^n}{\sum_{i=1}^N1/w^i}, ~~\forall n,t,
\] 
where $\ev_n$ is a unit vector with only its $n$th element being non-zero. With such a choice of $r_t$, any causal caching policy yields an expected utility  at most $CT/\sum_{n=1}^N(1/w^n)$, since:
\begin{align}
	\mean{\sum_{t=1}^T f_t\big(y_t\big)}
	&=\sum_{t=1}^T\sum_{n=1}^Nw^n\prob{r_t=\ev_n}y_t^n\label{eq:util}\\
	&\hspace{-0.2in}=\sum_{t=1}^T\frac{1}{\sum_n 1/w^n}\sum_{n=1}^Ny_t^n 
	\leq\frac{CT}{\sum_n 1/w^n}, \notag
\end{align}
To obtain a regret lower bound we show that a static policy with hindsight can exploit the knowledge of the sample path $\{r_t\}_{t=1}^T$ to  achieve a higher utility than \eqref{eq:util}. Specifically, defining $\nu^n_t$ the number of times file $n$ is requested in slots $1,\dots,t$, the best static policy will cache the $C$ files with highest products $w^n\nu^n_T$. In the following, we characterize how this compares against the average utility of \eqref{eq:util} by analyzing the order statistics of a Gaussian vector.

For i.i.d. requests we may rewrite the regret as the expected difference between the best static policy in hindsight and \eqref{eq:util}:
\begin{equation}\label{eq:regretIID}
R_T \geq \mean{ \max_{y\in\mathcal{Y}_1 }y^\top\sum_{t=1}^T w \odot r_t }- \frac{CT}{\sum_n 1/w^n},
\end{equation}
where $w \odot r_t=[w^1r^1_t, w^2r^2_t,...,w^Nr^N_t]^{\top}$ is the Hadamard product between the weights and request vector. Further, \eqref{eq:regretIID} can be rewritten as a function:
\begin{equation}
R_T\geq \mean{g_{N,C}(\overline{z}_T)} = \mean{\max_{b\in \stackrel{\circ}{\mathcal{Y}_1 }}\left[b^\top\overline{z}_T\right]},\notag
\end{equation}
\noindent where $\stackrel{\circ}{\mathcal{Y}_1}$ is the set of all integer caching configurations (thus, $g_{N,C}(\cdot)$ is the sum of the maximum $C$ elements of its argument); and the process $\overline{z}_T$ is the vector of utility obtained by each file after the first $T$ rounds centered around its mean:
\begin{align} \nonumber
\overline{z}_T & = \sum_{t=1}^{T}w \odot r_t - w \odot\frac{T}{\sum_{n=1}^N1/w^n}w^{-1} \\ \label{eq:centeredDemandVector}
& = \sum_{t=1}^{T}\left(z_t - \frac{1}{\sum_{n=1}^N1/w^n}\mathbf{1}_N\right) 
\end{align}
where $z_t = w \odot r_t$ are i.i.d. random vectors  with distribution 
\[ \mathbb{P}\left(z_t = w^i \ev_i\right) = \frac{1/w^i}{\sum_{n=1}^N1/w^n}, \forall t, \forall i,	\]
and, therefore, they have mean\footnote{Above we have used the notation $w^{-1} = \left[1/w^1, 1/w^2,...,1/w^N\right]^\top.$}:
\begin{equation}
 \mean{z_t}=\frac{1}{\sum_{n=1}^N1/w^n}\mathbf{1}_N. \notag
\end{equation} 
Key in our proof is the limiting behavior of $g_{N,C}(\overline{z}_T)$:
\begin{lemma}\label{lem:regretConvDistribution}
Let ${Z}$ be a Gaussian vector  $\Nc\left(\mathbf{0}, {\Sigmam}(w)\right)$, where ${\Sigmam}(w)$ is given in \eqref{eq:covarianceMatrix}, and ${Z}_{(i)}$ its $i-$th largest element. Then
\[
\frac{g_{N,C}(\overline{z}_T)}{\sqrt{T}} \xrightarrow[T\rightarrow\infty]{\text{distr.}} \sum_{i=1}^C{Z}_{(i)}.
\]
\end{lemma}  
\begin{IEEEproof} Observe that $\overline{z}_T$ is the sum of $T$ uniform i.i.d. zero-mean random vectors, where the covariance matrix can be calculated using \eqref{eq:centeredDemandVector}: ${\Sigmam}(w) =$
\begin{align}\nonumber
&=\mean{\left(z_1 - \frac{1}{\sum_{n=1}^N1/w^n}\mathbf{1}_N\right)\left(z_1 - \frac{1}{\sum_{n=1}^N1/w^n}\mathbf{1}_N\right)^\top}\\ \label{eq:covarianceMatrix}
&= \frac{1}{\sum_{n=1}^N1/w^n}\begin{cases}
w_i -\frac{1}{\sum_{n=1}^N1/w^n}, i=j\\
-\frac{1}{\sum_{n=1}^N1/w^n}, i\neq j
\end{cases},
\end{align}
where the second equality follows from the distribution of  $z_t$ and some calculations.\footnote{For the benefit of the reader, we note that $Z$ has no well-defined density (since ${\Sigmam}(w)$ is singular). For the proof, we only use its distribution.}
Due to the Central Limit Theorem: 
\begin{equation}\label{eq:convCenteredX}
\frac{\overline{z}_T}{\sqrt{T}} \xrightarrow[T\rightarrow\infty]{\text{distr.}} {Z}.
\end{equation}
Since $g_{N,C}(x)$ is continuous, \eqref{eq:convCenteredX}  and the Continuous Mapping Theorem \cite{Billingsley99} imply
\[
\frac{g_{N,C}\left({\overline{z}_T}\right)}{\sqrt{T}} = g_{N,C}\left(\frac{\overline{z}_T}{\sqrt{T}}\right) \xrightarrow[T\rightarrow\infty]{\text{distr.}} g_{N,C}\left({Z}\right),
\] 
and the proof is completed by noticing that $g_{N,C}(x)$ is the sum of the maximum $C$ elements of its argument.
\end{IEEEproof}

An immediate consequence of Lemma \ref{lem:regretConvDistribution} is that 
\begin{equation}
\frac{R_T}{\sqrt{T}}\!=\!\frac{\mean{g_{N,C}(\overline{z}_T)}}{\sqrt{T}}\xrightarrow{T\rightarrow\infty}\mean{\sum_{i=1}^C{Z}_{(i)}}\!=\!\sum_{i=1}^{C}\mean{{Z}_{(i)}} \notag
\end{equation}
and the first part of the Theorem is proved. 
	
To prove the second part, we remark that the RHS of the last equality is the expected sum of $C$ maximal elements of vector $Z$, and hence larger than the expected sum of any $C$ elements of $Z$. In particular, we will compare with the following: Fix a permutation $\bar{\phi}$ over all $N$ elements, partition the first $2C$ elements in pairs by combining 1-st with 2-nd, ..., $i$-th with $i$+1-th, $2C$-1-th with $2C$-th, and then from each pair choose the maximum element and return the sum. We then obtain: 
\begin{align}\nonumber
\mean{\sum_{i=1}^C{Z}_{(i)}} &\geq \mean{\sum_{i=1}^C\max\left[Z^{\bar{\phi}(2(i-1)+1)}, Z^{\bar{\phi}(2i)}\right]} \\ \nonumber
& = \sum_{i=1}^C\mean{\max\left[Z^{\bar{\phi}((2(i-1)+1)}, Z^{\bar{\phi}(2i)}\right]},
\end{align}
where the second step follows from the linearity of the expectation, and the expectation is taken over the marginal distribution of a vector with two elements of $Z$. We now focus on  $\max\left[Z^{k}, Z^{\ell}\right]$ for (any) two fixed $k,\ell$. We have that $(Z^{k}, Z^{\ell})^\top \sim \Nc\left(\mathbf{0}, \Sigmam(w^k, w^{\ell})\right)$, where $\Sigmam(w^k, w^{\ell})=$
\begin{align*}
&= \frac{1}{\sum_{n=1}^N1/w^n}\begin{bmatrix}
			w^k-\frac{1}{\sum_{n=1}^N1/w^n} & -\frac{1}{\sum_{n=1}^N1/w^n}\\
			-\frac{1}{\sum_{n=1}^N1/w^n} & w^{\ell}-\frac{1}{\sum_{n=1}^N1/w^n}
		\end{bmatrix}.
\end{align*}
From \cite{Clark1961} we then have:
\[
\mean{\max\left[Z^{k}, Z^{\ell}\right]} = \sqrt{\frac{1}{\sum_{n=1}^N1/w^n}}\frac{1}{\sqrt{2\pi}}\sqrt{w^{k}+w^{\ell}},
\]
therefore: 
\begin{equation}\label{eq:orderStatTransformed}
\mean{\sum_{i=1}^C{Z}_{(i)}} \geq \frac{1}{\sqrt{2\pi}}\frac{\sum_{i=1}^C\sqrt{w^{\bar{\phi}((2(i-1)+1)}+w^{\bar{\phi}(2i)}}}{\sqrt{\sum_{n=1}^N1/w^n}},
\end{equation}
for all $\bar{\phi}$. The result follows noticing that the  tightest bound is obtained by maximizing   \eqref{eq:orderStatTransformed}  over all permutations. \end{IEEEproof}

\begin{tcolorbox}[boxrule=0.7pt,arc=0.6em, colback=gray!3] 
	\begin{corollary}[Regret of Online Caching]\label{cor:2}
		Fix $C/N=\gamma$, $w^n=w$, for all $n$, and assume $C<N/2$, the regret of online caching satisfies:
		\[
		w\sqrt{\frac{\gamma}{\pi}} \sqrt{CT}\leq \min_{\sigma}R_T(\sigma) \leq w\sqrt{2}\sqrt{CT}~~~\text{as}~T\to\infty.
		\]
	\end{corollary}
\end{tcolorbox}
\noindent Corollary \ref{cor:2} follows from  Corollary \ref{cor:lbound1} and Theorem \ref{th:2}. We conclude that omitting $\sqrt{2\pi/\gamma}$, which i s amortized by $T$, BSCA achieves the best possible learning rate for the one cache problem.\footnote{We note that in the special case where the term $\gamma=C/N$ diminishes as the time horizon $T$ increases, we do not obtain matching bounds and the question of what is the fastest learning policy is open.}

%% file: extensions_R1.tex
\section{Model Generality and  Extensions}\label{sec:extensions}

The proposed model and algorithm can be used to solve different instances of (OCP) and other problems that have similar structure. We discuss some representative cases next. 

\begin{figure}
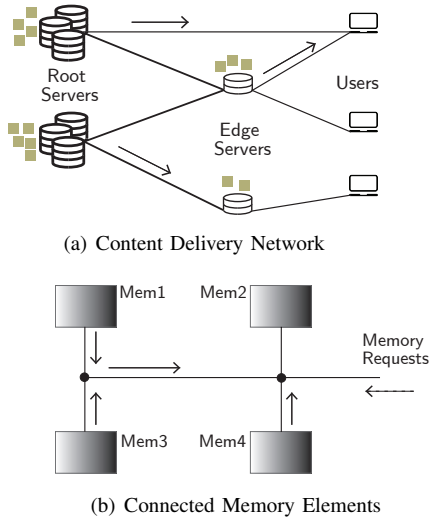

	\centering
	\subfigure[Content Delivery Network]{
		\centering
		\includegraphics[width=1.98in]{intro-model2}\label{Fig:cdn}
	}\,\,\,\,\,\,\,\,\,\,\,\,\,\,\,\,\,\,\,\,
		\subfigure[Connected Memory Elements]{
		\centering
	\includegraphics[width=1.98in]{disag-example}\label{Fig:disag}}
	\caption{\textbf{Bipartite model for general networks}. (a) Content delivery network with root and edge servers, and uncapacitated links. Each path can be modeled as a super-link in a bipartite graph of (root or edge) servers and end-users. (b) Network of connected memory elements in a disaggregated server system. }
	\label{Fig:bipartite}
\end{figure}

\textbf{General Graphs}. An arbitrary network can be modeled with a bipartite graph if it does not have hard link capacity constraints or load-dependent routing costs. This is achieved by calculating the best path connecting a user to any reachable cache, and using this path cost as the link cost in the bipartite model. With this transformation our analysis applies to a very broad class of caching nework problems. Two such examples are Content Delivery Networks (CDN) if their links have large enough capacities, Fig. \ref{Fig:bipartite}(a); and multi-memory paging systems appearing in data centers and {disaggregated} server architectures \cite{dredbox}, \cite{disag}, Fig. \ref{Fig:bipartite}(b). 

\textbf{Dynamic Networks}. The links connecting user locations to edge caches might be heterogeneous, introducing different costs in terms of transmission power or delay. This aspect can be modeled through the utility functions in order to reflect that users would prefer to obtain the files over low-cost links. We can redefine the utility parameters as a product of a cache-related benefit and a link-related benefit, i.e., $w^{n,i,j}\!=\!w_{cach}^{n,j}\cdot w_{rout}^{i,j}$, and then apply BSCA. The steps are identical as in Algorithm 1, and the only change is in the constants of the regret bound. Moreover, our model can capture the case where utility parameters change over time, $w_{t}^{n,i,j}$, e.g., due to link costs changing or users moving from one location to another. This extension, indirectly, allows to include mobility models in our analysis. BSCA can handle this effect by replacing the step in line 5 where in each slot we need now to observe both the submitted request $r_{t}^{\hat{n}, \hat{i}}$ and the current utility vector $w_{t}$. This generalization does not affect the regret which already includes the maximum utility distance.

\textbf{Reconfiguration Costs}. Finally, an important case arises when there is cost for prefetching the files over the backhaul SBS links. First, note that BSCA might select to reconfigure the caches in a slot $t$ ($y_t\neq y_{t-1}$) even if the requested files were already available, if this update is expected to improve the total utility. However, such changes induce bandwidth cost and therefore the question ``\emph{when should a file be prefetched?}'' arises naturally. Unlike other policies that make also such proactive updates, e.g., the LRU-ALL policy \cite{giovanidis-mLRU}, BSCA can take into account these costs and reconfigure when they are smaller than the expected benefits. In detail, if we denote with $c_{n,j}$ the cost for transferring a file unit from the origin servers to cache $j$, we can define the utility-cost function:
\begin{equation}
	J_t(y_t, y_{t-1})=f_{t}(y_t)  - \sum_{n\in\mathcal{N}}\sum_{j\in\mathcal{J}}c_{n,j}\max\{y_{t}^{n,j}- y_{t-1}^{n,j},0\}\nonumber
\end{equation}
where $f_t(y_t)$ is given in \eqref{eq:biput}, and the convex $\max$ operator ensures that we pay cost whenever we increase the cached chunks of a file $n$ at a cache $j$ (but not when we evict data). Function $J_t$ is concave and hence BSCA can be employed. Namely, it suffices to use the supergradient $q_t$ for $J_t(\cdot)$ instead of that for $f_t(\cdot)$. Denoting $q_t$ the supergradient of $J_t$ and using basic subgradient algebra we can write $q_t=g_t + h_t$, where
\begin{align}\label{eq:supergrad2}
	h_{t}^{\hat{n},j}=\left\{\begin{array}{ll}
		-c_{ \hat{n},j }, & \text{if  }\,y_{t}^{\hat{n},j} - y_{t-1}^{\hat{n},j}>0 \\
		0 & \text{otherwise}
	\end{array}\right.
\end{align}
which is calculated for each request $r_{t}^{ \hat{n}, j }$ and cache $j$. The policy's learning rate is not affected by this change, and we only need to redefine $K$ by adding the maximum value of $h$. This makes our policy suitable for placement problems beyond caching, e.g., costly deployment of in-network services.

%% file: simulations_R1.tex
\section{Performance Evaluation}\label{sec:performance}

We evaluate the performance of several policies in terms of hit rate and accrued utility, using different request sequences. We see that although BSCA might not always achieve the highest performance, it is in every case close to the best-performaning policy, something that is not true for any other benchmark. We begin with the single cache problem which highlights some interesting features of BSCA, such as the hybrid recency/frequency criterion it uses to cache files. We then compare BSCA with state-of-the-art reactive policies in caching networks, namely multi-LRU \cite{giovanidis-mLRU} and q-LRU \cite{leonardi-implicit}. We find that BSCA outperforms these policies and creates, asymptotically, average utility equal to that of the {best static cache configuration}.

\begin{figure*}[!t]
	\centering
	\subfigure[\footnotesize{CDN aggregation (IRM)}] 
	{\includegraphics[width=1.6in]
		{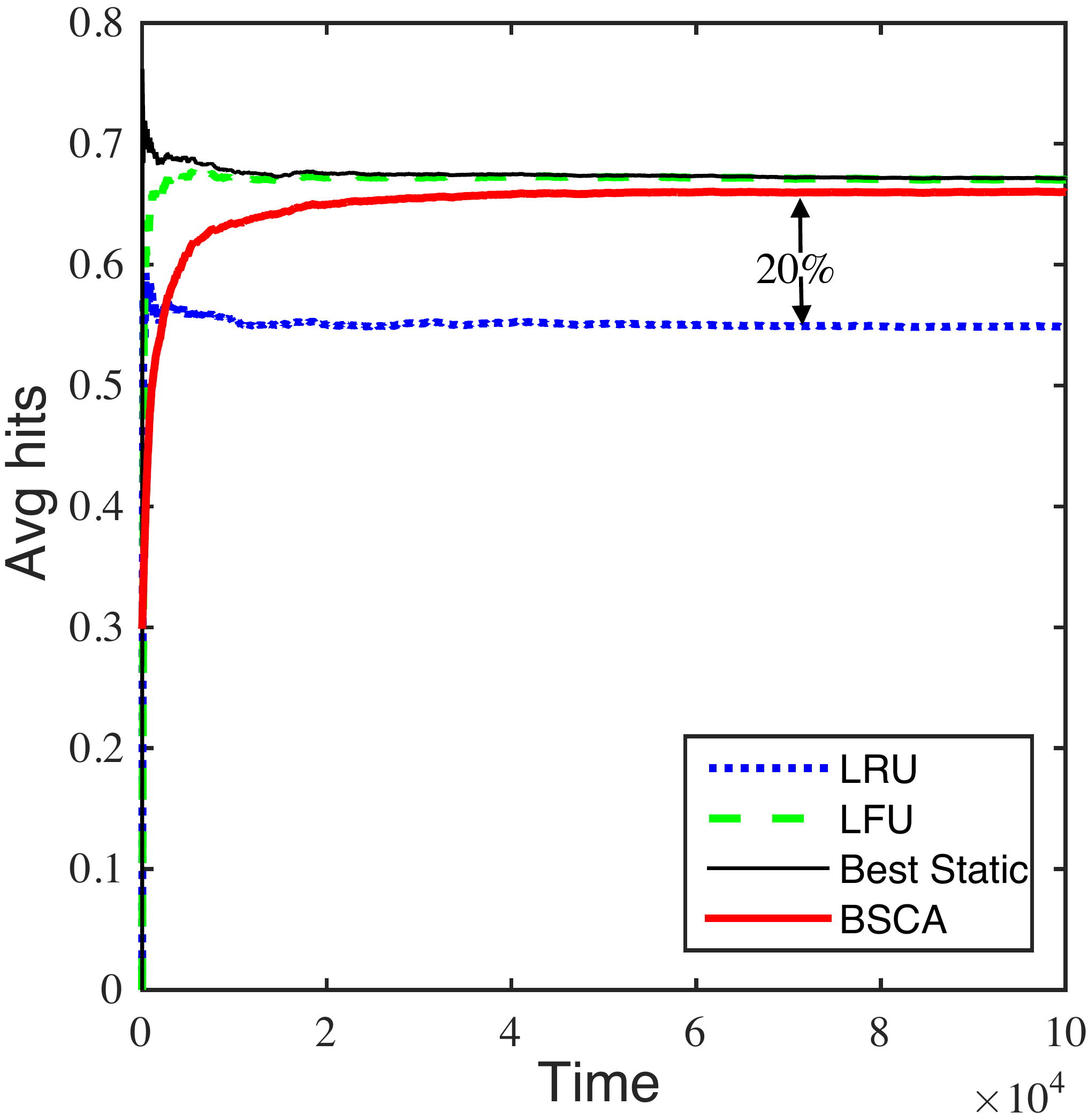}
		\label{Fig:Fig1}}
	\subfigure[\footnotesize{YouTube videos \cite{snm}}]{\includegraphics[width=1.6in]
		{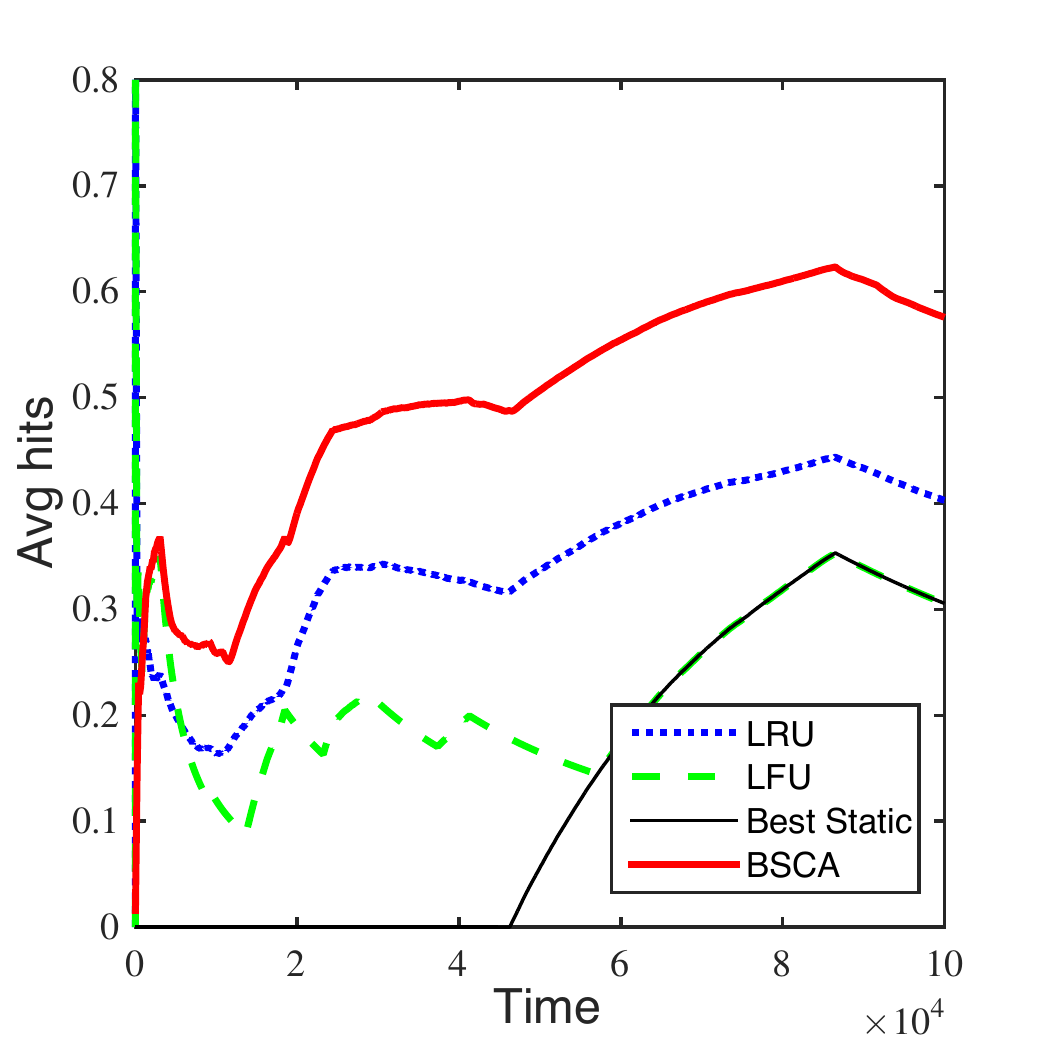}
		\label{Fig:Fig2}}
	\subfigure[\footnotesize{Web browsing (1) \cite{Kurose08}}]
	{\includegraphics[width=1.6in]
	{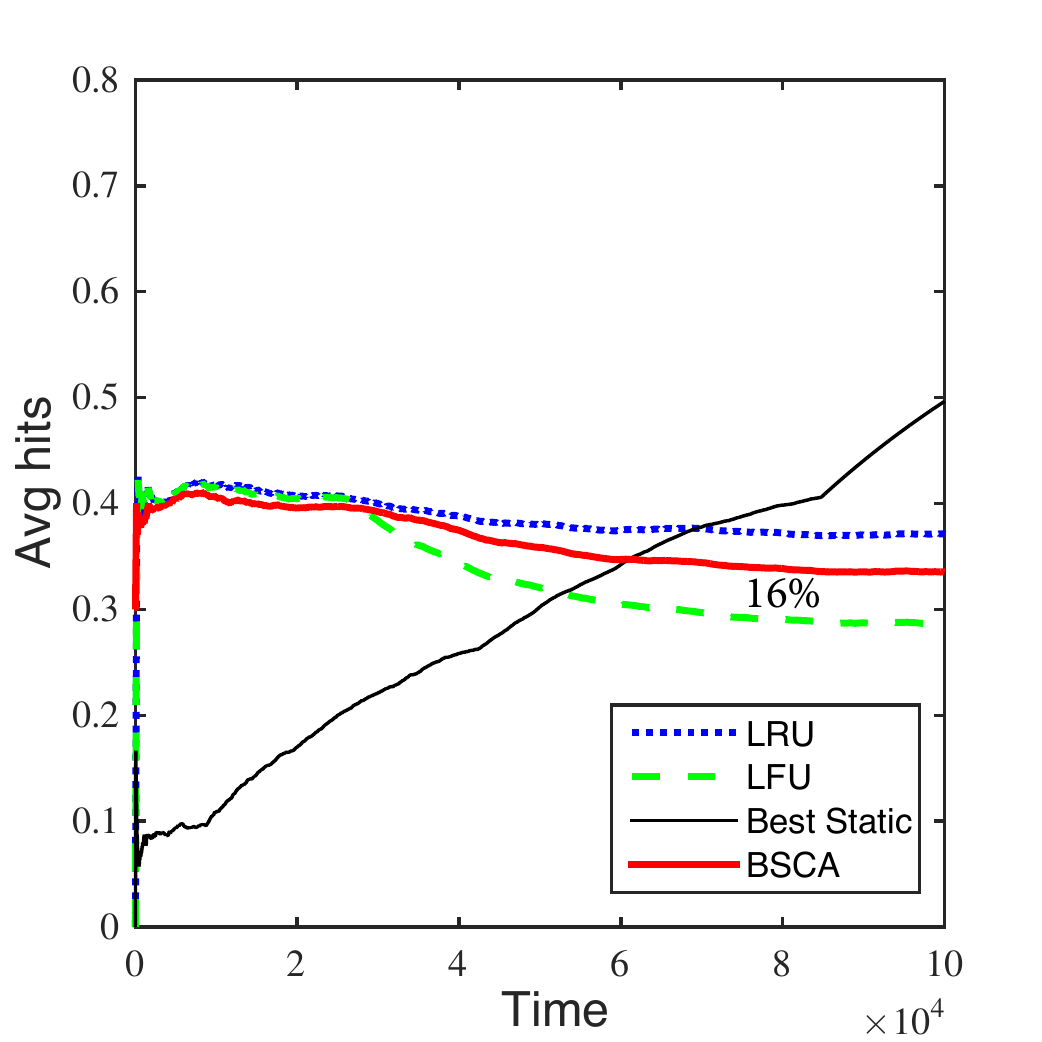}
	\label{Fig:Fig3}}
	\subfigure[\footnotesize{Web browsing (2) \cite{Kurose08}}]
	{\includegraphics[width=1.6in]
	{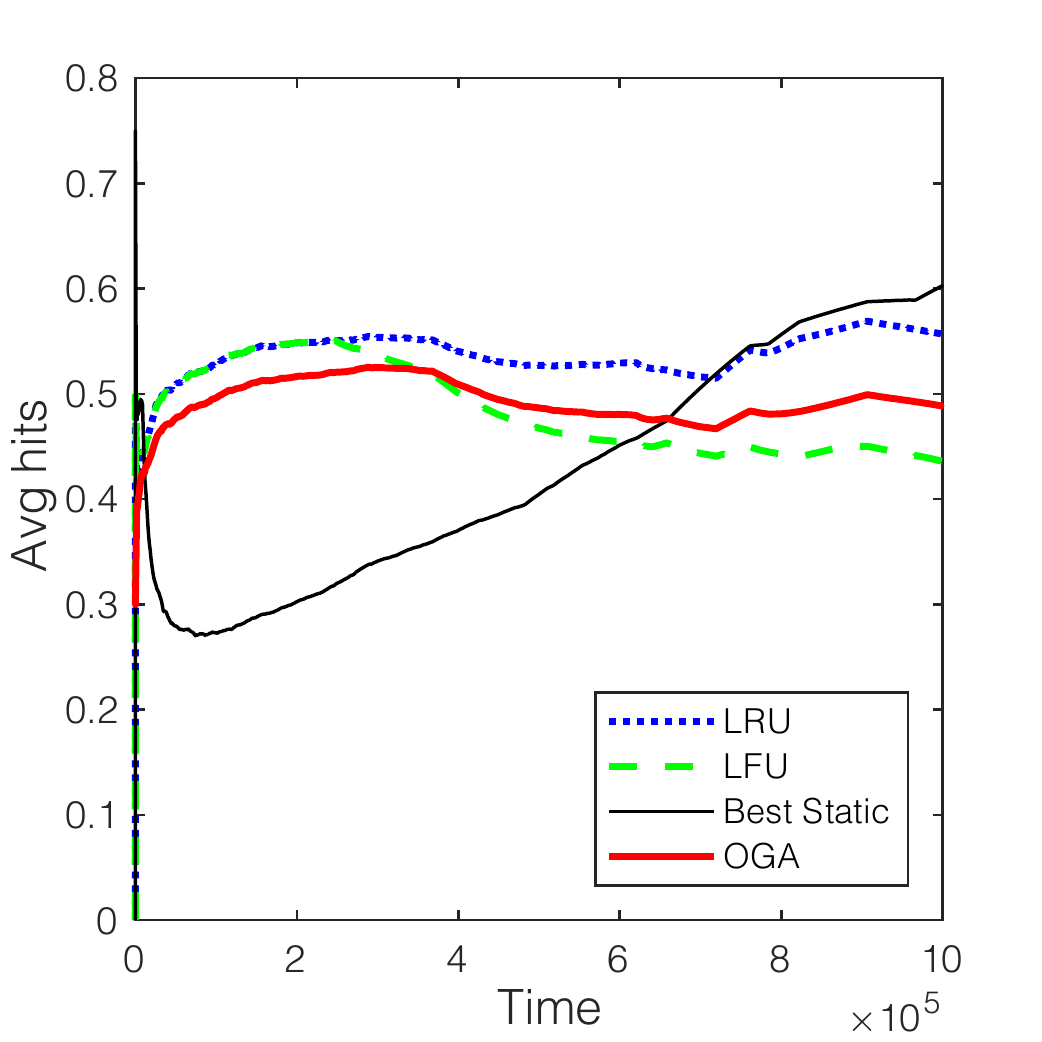}
	\label{Fig:Fig4}}	
	\caption{\textbf{Average Number of Hits for BSCA in one cache}: We consider the following request models: (a) i.i.d. Zipf \cite{fricker2012impact}; (b) Poisson Shot Noise \cite{snm}; (c) web browsing dataset \cite{Kurose08}; (d) web browsing dataset \cite{Kurose08} for 10 times longer time horizon.} 
	\label{Fig:perf2}
\end{figure*}

\begin{figure}[!t]
\centering
\includegraphics[width=1.94in,height=1.94in]{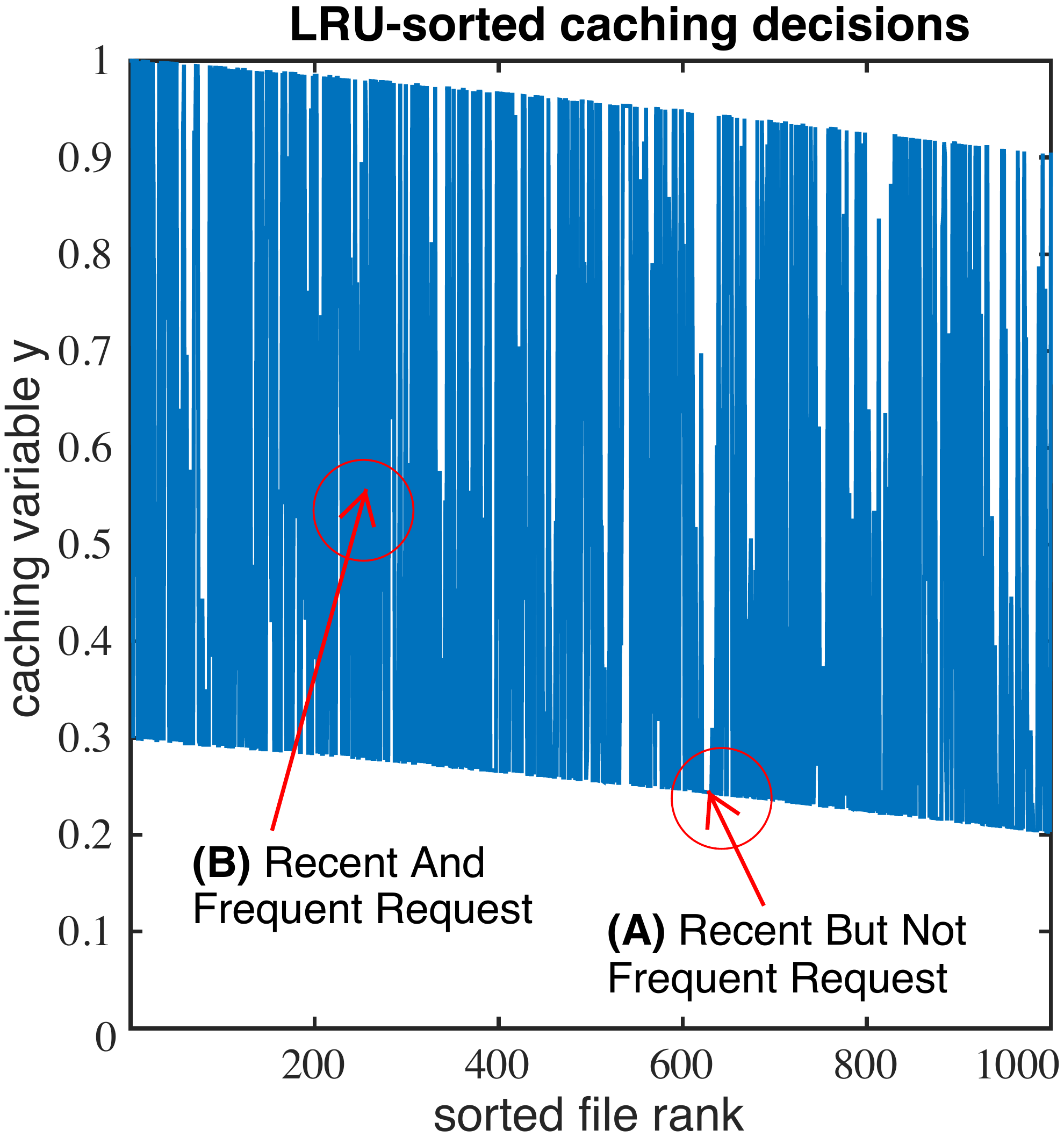}
\label{Fig:lfu-lru}
\caption{\textbf{Comparison of BSCA with LRU, $N\!=\!10^4$, $C\!=\!10^3$}: For each file in the LRU cache, we show the corresponding values of the BSCA caching variable. This reveals how much of this file BSCA would have stored, and how it differs from the decisions of LRU.} 
\end{figure}

\subsection{Single Cache Scenarios}

In these experiments we use a cache with capacity equal to $30\%$ of the (observed) library in each scenario, unless otherwise stated. We test different request models: \emph{(a)} an i.i.d. Zipf model that represents requests in a CDN \cite{fricker2012impact}; \emph{(b)} a Poisson shot noise model that represents ephemeral YouTube video requests \cite{snm}; and \emph{(c)} a dataset from \cite{Kurose08} {with actual web browsing requests} at a university campus. This latter model draws files, naturally, from a larger file library. For cases \emph{(a)} -- \emph{(c)} we use $T=10^5$ requests, and for case \emph{(d)} $T=10^6$ requests. The results are shown in Figure \ref{Fig:perf2} where we plot the evolution of the cache hit ratio (i.e., hits versus total current requests) for BSCA, LRU, LFU and the best static configuration. It is important to emphasize that the latter is decided using the entire time horizon $T$, and hence might even yield zero hits for certain time intervals.

We observe that the performance of BSCA is very close to the best among LFU and LRU for each scenario, after we allow for some adaptation time. Namely, the relative gain (in terms of hits) compared to the second best policy is as high as 20$\%$ over LRU (Fig. \ref{Fig:Fig1}) and up to 16$\%$ over LFU (Fig. \ref{Fig:Fig3}), while BSCA outperforms both in case \emph{(b)}.\footnote{The percentages are the relative gains in terms of cache hit ratio, i.e., we divide the difference of BSCA hits minus LRU hits, with this latter quantity.} Note that it is even possible for BSCA and the other reactive policies to outperform the static benchmark.

\begin{figure*}
	\centering
	\subfigure[Example of bipartite caching network.]
	{
		\includegraphics[width=1.8in]{bipartite-example}
		\label{Fig:bipartite-example}
	}
	\subfigure[Utility of BSCA \& Competitors]
	{
		\includegraphics[width=1.8in, height=1.7in]{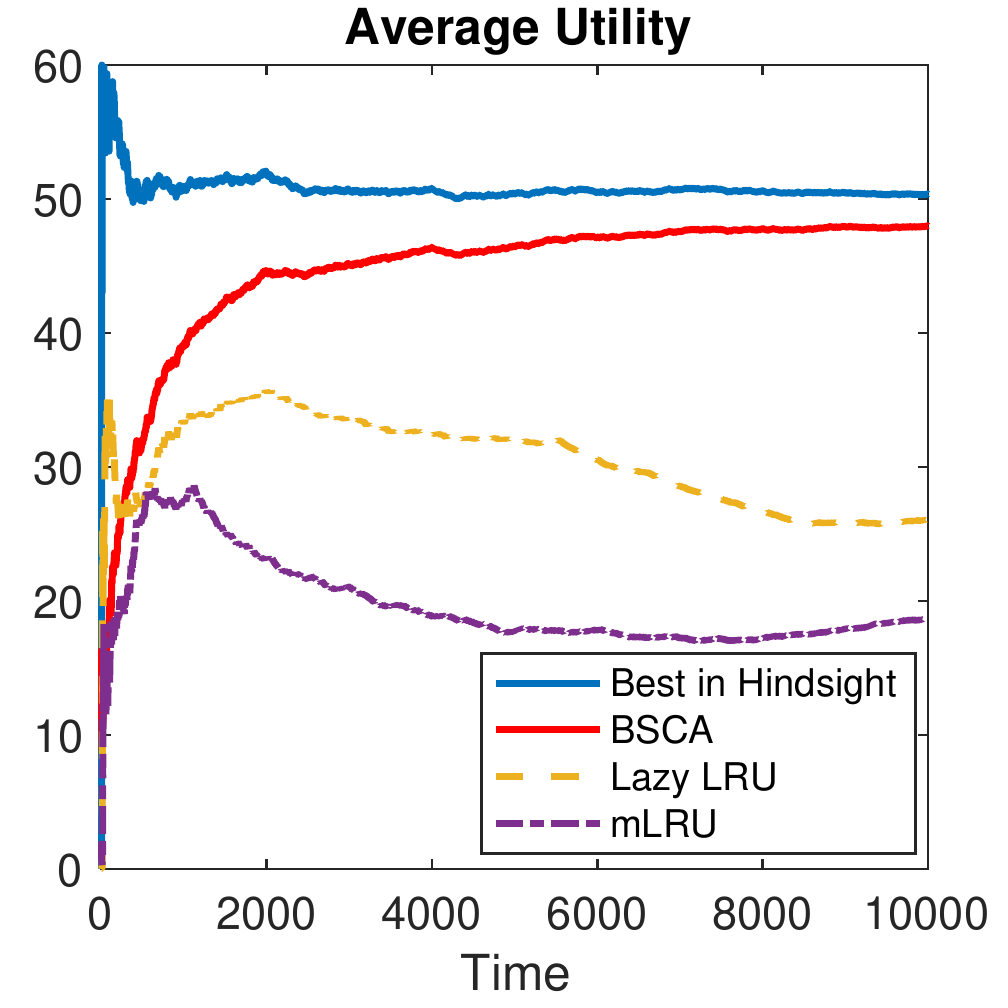}
		\label{Fig:bipartite-utility}
	}
	\subfigure[Utility of BSCA \& Competitors]
	{
		\includegraphics[width=1.8in, height=1.7in]{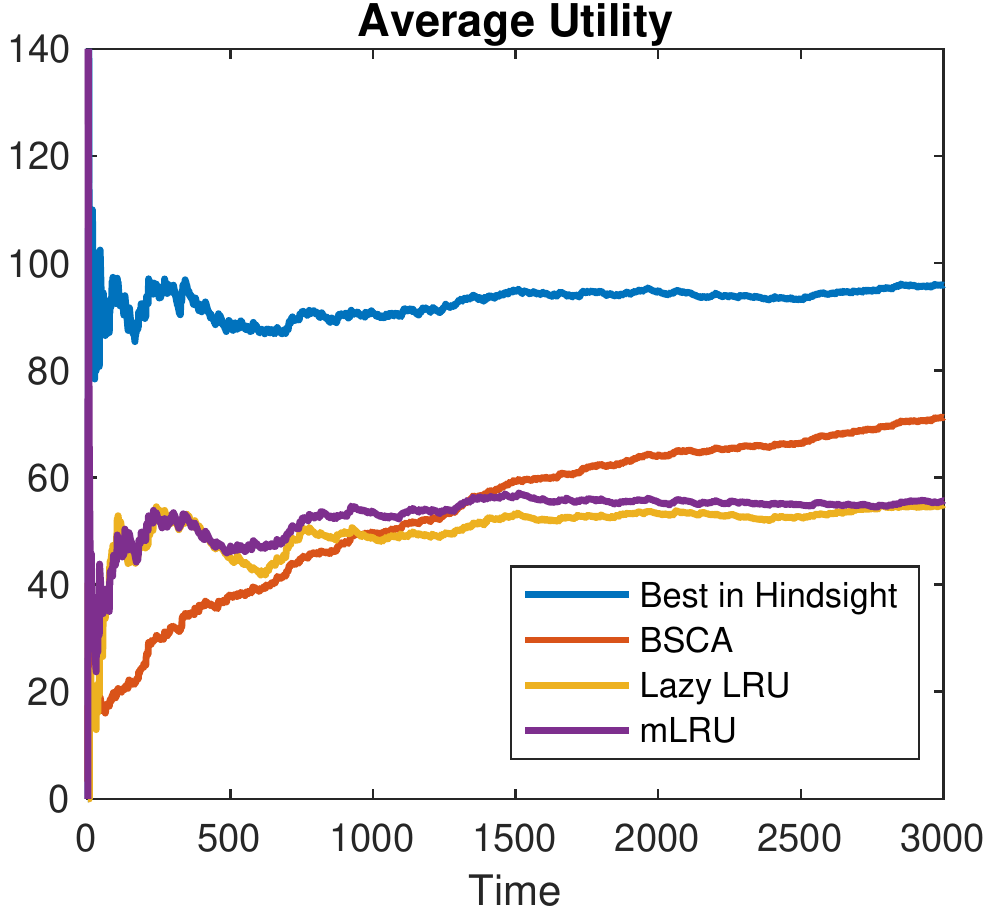}
		\label{Fig:bipartite-utility2}
	}
	\caption{\textbf{BSCA for a bipartite network}. (a) User $2$ is connected to caches $j=1, 2$; a request $r_{t}^{n,2}$ arrives, and serving it from cache $1$ yields $w^{n,2,1}$. (b) The average utility of BSCA and competitor policies for this network. (c) New comparison for a network with $J=6$ caches, $I=14$ locations, and utilities that are uniformly random in $[0,100]$.} 
	\label{Fig:bipartite-sims}		
\end{figure*}


Taking a closer look in these experiments, we can identify the source of the remarkable agility of BSCA. First, recall that LFU calculates the request frequency of each item and evicts the one that is (up to this slot) less frequently requested. On the other hand, LRU keeps track of how recently an item was requested in the past, and evicts the one that was least recently requested. BSCA bears similarities to both policies but uses a Euclidean regularizer (in the projection step) which yields smoother --- and, it turns out, wiser --- decisions.\footnote{The regularization becomes clear if one sees that our subgradient iteration $y_{t+1}=\Pi_{\mathcal{Y}}(y_t +\eta_tg_t)$ can be viewed as linearization of the proximal algorithm \cite{beck-mirror}, i.e., $y_{t+1}=\arg\min_{y\in\mathcal{Y}} \{ \langle y,g_t \rangle + \frac{1}{2\eta_t}\|y-y_t \|^2  \}$.}. Namely, it decides which chunks to cache based on the recently submitted requests (recall that for 1 cache it is $g_t\!=\!r^{ \hat{n},\hat{i}}w^{ \hat{n}, \hat{i} }$) resembling LRU, but also makes these decisions gradually, requiring multiple requests for each file, similarly to LFU. In other words, BSCA can be viewed as a normalized version of utility-LFU policy, where the frequency-based eviction decisions are smoothened; and similarly considers the recency as LRU but reacts with inertia.

This is demonstrated in Fig. 5 which shows an example of the LRU decisions ($C\!=\!1$K most recently used) and the respective BSCA decisions for these files. This reveals that the two policies take strongly correlated decisions, but BSCA additionally ``remembers'' the frequency of requests. For instance, point (A) represents a file $n_A$ that was requested recently but infrequently overall, and hence BSCA decreases the value of $y^{n_A}$ just like LFU. On the other hand, point (B) represents a file $n_B$ that was both recently requested and frequently, and hence BSCA sets large value for $y^{n_B}$.

\subsection{Bipartite Caching Networks}

Next we consider a bipartite graph with 3 caches of size $C=10$ and 4 user locations, Fig.~\ref{Fig:bipartite-example}. The utility vector is $w^{n}=(1,2,100), \forall n$, hence an efficient policy needs to place popular files on cache $3$. The network is fed with stationary Zipf requests from a library of $N=100$ files, and each request arrives at a user location that is selected uniformly at random. We stress that we have chose a small value for $N$ only to facilitate the calculation of the hindsight policy. We compare BSCA to the best static configuration, and state-of-the-art reactive policies: \emph{(i)} the multi-LRU policy proposed in \cite{giovanidis-mLRU} where a request is routed to a given cache (e.g., the closest) which is updated based on the LRU rule; and \emph{(ii)} and the $q$-LRU policy with the ``lazy'' rule \cite{leonardi-implicit} for $q=1$, which works as the multi-LRU but updates the cache only if the file is not in any other reachable cache.

Fig. \ref{Fig:bipartite-utility} presents the results for the highly asymmetric (in terms of utility) network of Fig. \ref{Fig:bipartite-example}, while Fig. \ref{Fig:bipartite-utility2} presents a similar experiment for a larger network with symmetric utilities. Please note that here, unlike Fig. \ref{Fig:perf2}, we calculate the optimal static configuration for the period up to each slot $t$ and compare its performance with that of BSCA -- this allows us to examine how the quantity $R_T/T$ evolves with $T$, and observe in practice how BSCA learns. Indeed, we see that BSCA converges to the best static hindsight policy in both cases, which verifies that it is a universal no-regret policy. In other words, BSCA gradually learns which files are popular and increases their placement at the high utility caches. For the first experiment in Fig. \ref{Fig:bipartite-utility}, the second best policy is lazy-LRU which is outperformed by BSCA by $45.8\%$. On the other hand, both lazy-LRU and mLRU have comparable performance for the experiment in Fig. \ref{Fig:bipartite-utility2} and, interestingly, BSCA has a lower performance for the first 1000 slots but quickly adapts to the requests and outperforms its competitors.

	


%% file: conclusions_R1.tex
\section{Conclusions}\label{sec:conclusions}
 
The seminal femtocaching proposal \cite{femtocaching} initiated a fascinating research thread on wireless edge caching that extends beyond content delivery, to deployment of  services and to edge computing. One limitation of this idea is that it presumes the existence of a static and known popularity model for the requests. Previously proposed solutions either try to estimate the (assumed) fixed popularity or fit intricate non-stationary models to data. Taking a fundamentally different path, we design here an online network mechanism that adapts the caching and routing decisions to any request pattern, even one that is designed by an adversary, and converges to the optimal static performance. To achieve this, we employed online convex optimization and developed a learning algorithm that is simple, fast, and fully-embedded to the network operation. The regret of our policy is sublinear on time and independent of the content catalog size. The algorithm achieves the best possible regret for the single-cache case, and we have also established a link between this OCO-based policy and the classical LRU/LFU policies. Trace-based experiments demonstrate the effectiveness of our approach in different cases. 

This work brings together the theory of online convex optimization and (wireless or wired) caching networks, and paves the road for the principled design of dynamic caching and routing policies. Exciting future research directions include, among others, the extension of these ideas to non-bipartite graphs, to systems that allow dynamic storage placement or cache rescaling, and to incorporate resource allocation decisions such as transmission power or link scheduling. 